\documentclass[10pt,sigconf]{acmart}
\usepackage[linesnumbered,ruled]{algorithm2e}
\usepackage{amsmath,epsfig,bm,graphicx}
\usepackage{empheq}
\usepackage{cleveref}
\usepackage{makecell}
\usepackage{booktabs}
\usepackage{multirow}
\usepackage[section]{placeins}
\graphicspath{{figs/}}

\def\BibTeX{{\rm B\kern-.05em{\sc i\kern-.025em b}\kern-.08emT\kern-.1667em\lower.7ex\hbox{E}\kern-.125emX}}
    
\copyrightyear{2019}
\acmYear{2019}
\setcopyright{acmlicensed}
\author{Renzhi Wu}
\email{wurenzhi@foxmail.com}
\orcid{0000-0002-9144-8999}
\affiliation{%
  \institution{AGT International}
  \city{Darmstadt}
  \state{Germany}
}

\author{Sergey Sukhanov}
\email{ssukhanov@agtinternational.com}
\affiliation{%
  \institution{AGT International}
  \city{Darmstadt}
  \state{Germany}
}

\author{Christian Debes}
\email{cdebes@agtinternational.com}
\affiliation{%
  \institution{AGT International}
  \city{Darmstadt}
  \state{Germany}
}
\begin{document}
\fancyhead{}
\title{Real Time Pattern Matching with Dynamic Normalization}

\begin{abstract}
Pattern matching in time series data streams is considered to be an essential data mining problem that still stays challenging for many practical scenarios.
Different factors such as noise, varying amplitude scale or shift, signal stretches or shrinks in time are all leading to performance degradation of many existing pattern matching algorithms.
In this paper, we introduce a dynamic z-normalization mechanism allowing for proper signal scaling even under significant time and amplitude distortions. 
Based on that, we further propose a Dynamic Time Warping-based real-time pattern matching method to recover hidden patterns that can be distorted in both time and amplitude.
We evaluate our proposed method on synthetic and real-world scenarios under realistic conditions demonstrating its high operational characteristics comparing to other state-of-the-art pattern matching methods.

\end{abstract}


%\ccsdesc[500]{Information systems~Data mining}


\maketitle

\section{Introduction}
\label{sec:intro}
Pattern matching in data streams (also known as similarity search, subsequence matching or stream monitoring) has recently emerged as an important task for the data mining community with application in many other domains including finance, industry, health care, networks, etc.~\cite{fu2011review,DTW_million,PAMAP}.
Accurate real time subsequence matching from a data stream might allow to prevent damages to industrial equipment or infrastructure, timely recognize development of severe health complications or avoid traffic collision~\cite{UCRArchive2018}.
The data stream signals in such applications often suffer from the presence of noise, potential time distortions (due to the variance in sampling rate or just the nature of the underlinying process) and varying amplitude scale~\cite{keogh2003need}. 
Subsequence matching under such conditions is considered to be a difficult task offering many challenges to be addressed.

Dynamic Time Warping (DTW) has emerged as a natural way to compare a pair of time series sequences and later became a primary tool for solving pattern matching problems~\cite{fu2011review}.
Initially proposed for speech recognition tasks~\cite{kruskalsymmetric}, DTW was sucessfully applied to many other domains proving to be a powerful technique for sequence comparison and retrival~\cite{berndt1994using, rath2003word, muller2007dynamic}: numerous experiments have demonstrated its superiority over other methods~\cite{ding_trajcevski_scheuermann_wang_keogh_2008,DTW_million}. 
Formally, DTW is a dissimilarity measure that allows to find the best alignment between two sequences reporting their degree of mismatch. 
Despite the fact that, strictly, DTW is not a distance metric it is nevertheless used by many algorithms in order to align and compare a pair of time series signals, potentially of different lengths. Currently, pattern discovery has become one of the domains where DTW is extensively and successfully utilized.

One of the pattern matching approach utilizing DTW is the SPRING algorithm~\cite{sakurai_faloutsos_yamamuro_2007} that was initially proposed in order to accurately solve pattern matching problem in real time. 
SPRING is built on top of two major ideas: star-padding that reduces the time and space complexity to linear with respect to the data stream size guaranteeing that the minimum distance is obtained; and subsequence time warping matrix (STWM) that is an extension of a warping matrix natively utilized by DTW.
%and each cell in the matrix stores the starting position and the distance so far for each candidate subsequence. 
%Peng et al.~\cite{peng_liang_yan_weihong_shuqiang_2008} later proposed a pruning technique to speed up SPRING while Niennattrakul et al.~\cite{niennattrakul_wanichsan_ratanamahatana_2010} improved the matching performance of this approach proposing subsequence length constraints. 
Later, Gong et al.~\cite{gong_si_fong_mohammed_2014, gong_fong_si_2018} introduced a normalization-supported SPRING (NSPRING) that performs normalization of streaming data by considering mean and standard deviation of a predefined number of subsequent data samples. 
Giao et al.~\cite{ISPRING} proposed ISPRING that integrates min-max normalization into SPRING by considering the minimum and maximum values on a fixed length monitoring window.

Another DTW-based pattern matching approach called UCR-DTW was proposed in the UCR-suite~\cite{DTW_million} to achieve normalized subsequence discovery on large scale time series data by efficient pruning techniques. 
Similar to NSPRING and ISPRING, UCR-DTW employs a fixed length sliding window in order to normalize the streaming data and discover patterns hidden in a stream.

Normalizing streaming data in a pattern matching problem is considered to be an important~\cite{DTW_million, keogh2003need} though challenging task due to the real-time nature of the data and unknown position of hidden patterns in the stream. 
An appropriate normalization procedure usually allows to bring the query and the streaming subsequence to comparable ranges letting pattern matching algorithms focus on the structural similarities rather than just on the amplitude levels. 
The predominant practice for normalizing data stream is to employ normalization approaches such as min-max~\cite{ogasawara2010adaptive} or z-normalization~\cite{tan_steinbach_kumar_2014} with a fixed length sliding window %(might be named differently or implicitly) 
on the data stream~\cite{gong_si_fong_mohammed_2014,ISPRING,DTW_million}. Generally, fixed length sliding window-based normalization methods often cause unwanted structural disturbances, compromising consequent DTW by damaging its natural robustness to time distortion.
The reason is that the potential variation of the length of the hidden-in-stream pattern is an intrinsic assumption of DTW (stretches of signals are possible). 
%Additionally, fixed-length normalization windows can also cause a time delay that might be critical in applications where decisions have to be made rapidly.

Another critical problem is time distortion known as uniform scaling that is a global stretching or shrinking of time series in the time direction (t axis)~\cite{keogh2004indexing}. 
%The naive way to handle uniform scaling is to iteratively scale subsequences to a predefined length interval before matching~\cite{aach2001aligning, argyros2003efficient}, which is computationally inefficient. 
The naive solution to takle this %way to handle uniform scaling
is to iteratively scale subsequences of different lengths to the length of the template sequence before matching~\cite{aach2001aligning, argyros2003efficient}. However, this is computationally inefficient and hardly tractable in practical applications. 
To speed up calculation under DTW distance, several lower bounding techniques were proposed~\cite{fu2008scaling, shen2017searching, shen_chen_keogh_jin_2018}, however, none of these lower bounds support signal normalization while extending them for that would increase the time complexity tremendously~\cite{shen_chen_keogh_jin_2018}. The current state-of-the-art that supports both normalization and uniform scaling is UCR-US (part of UCR-suite~\cite{DTW_million_journal}). However, instead of DTW, UCR-US employs Euclidean distance that limits its practical application.
%For example, extending the state-of-the-art lower bound~\cite{shen_chen_keogh_jin_2018} to supporting normalization on the corresponding scaling interval of a scaling factor causes the time complexity for computing the lower-bound value alone to be $O(nl^2m^2\text{log}(m))$ (where $n$ and $m$ are the length of the data sequence and query sequence; $l$ is the maximum scaling factor), which is already comparable to the time complexity of a brute-force pattern search $O(nl^2m^3)$ and would be only slightly better than the brute-force search after adding the computations that the lower bound is unable to prune.
%Besides, all of these lower bounds require a predefined maximum scaling factor which is unknown in practice, and a loose scaling factor would increase the computational burden dramatically as the time complexity is proportional to the square of the maximum scaling factor.

To address above mentioned limitations of the fixed window length-based normalization methods and the effect of uniform scaling in pattern matching problem, in this paper, we introduce a dynamic z-normalization mechanism that progressively scales incoming samples of the evolving data stream.
By utilizing dynamic z-normalization and adopting STWM concept~\cite{sakurai_faloutsos_yamamuro_2007} we propose a DTW-based pattern matching approach that is able to accurately report discovered subsequences that contain possible amplitude distortions in real-time.
Thanks to the additive property of the dynamic z-normalization, the proposed pattern matching approach is able to discover hidden patterns under relatively large time distortions without the need of any predefined scaling parameters and it is %at least $l^2m\text{log}(m)$ times 
faster than the state-of-the-art lower bounding techniques (extended to supporting normalization).% in both the worst case and the amortized case. 
The implementation of the proposed method can be found at~\cite{wurenzhi2020Apr}.

The paper is structured as follows. In Section~\ref{sec:problem_formulation} we formulate the problem of pattern matching reviewing the traditional fixed length window z-normalization method. In Section~\ref{sec:proposed_method}, we introduce a dynamic z-normalization mechanism and further based on that we propose a DTW-based pattern matching approach that provides accurate and instant results for stream monitoring tasks. We evaluate the proposed pattern matching method in Section~\ref{sec:experiments} on synthetic and real-world datasets reporting its operational performance and concluding our paper in Section~\ref{sec:conclusions}.

%On data that doesn't need normalization, SPRING algorithm naturally handles a certain amount of uniform scaling with the star padding technique.
%The sliding window approach is 
%For data that requires normalization, the iterative scaling process is combined with fixed length sliding window and pruning lower bounds.
%That is to scale subsequences to a predefined length interval, normalize on each length and speed up calculation with pruning. 

%However, all of these only provides very loose lower bounds; the state-of-the-art lower bound takes $2.5$ h to query on the REFIT dataset~\cite{hargreaves2015smart} that contains $7,633,070$ data point with a maximum scaling factor of $2.5$ ~\cite{shen2017searching}.

\section{Problem formulation}
\label{sec:problem_formulation}
Consider a varying in time data stream $S$ to be a semi-infinite ordered set $\{s_0,s_1,s_2,\dots,s_t\}$, where $s_t$ is the most recent value (at the current time tick $t$). 
Let $S[t_b{:}t_e] = \{s_{t_b},s_{t_b+1},\dots,s_{t_e}\}$ be a subsequence of $S$ beginning with $s_{t_b}$ and ending with $s_{t_e}$. 
The general objective of a pattern matching problem is formulated as follows: given a query (template) sequence $Q = \{q_0,q_1,\dots,q_{m-1}\}$ with a fixed length $m$, find the subsequences of $S$ that are similar to $Q$. 
Similarity is usually replaced by the distance measure that has to provide a small value between $Q$ and $S[t_b{:}t_e]$ in case of their match.
We denote the distance between $Q$ and $S[t_b{:}t_e]$ as $D(S[t_b{:}t_e],Q)$.
In this paper, we consider the three following formal problems of pattern matching:

%We denote the distance between $Q$ and $S[t_b{:}t_e]$ as $D(S[t_b{:}t_e],Q)$. 
%This general problem formulation can be further decomposed into three following types of problems depending on the task at hand~\cite{sakurai_faloutsos_yamamuro_2007}.
\textit{Problem 1: real-time monitoring.} Report the subsequence $S[t_b{:}t_e]$ in real time (before time tick $t_e+1$) once $D(S[t_b{:}t_e],Q) \leq \epsilon$; where $\epsilon$ is a predefined threshold. 
Problem 1 can be further extended to two derivative problems depending on the task at hand~\cite{sakurai_faloutsos_yamamuro_2007}. 

\textit{Problem 2: disjoint query.} %Either when $S$ is pre-recorded or it is updating online, 
Find all subsequences $S[t_b{:}t_e]$ that satisfy two conditions: 

a) $D(S[t_b{:}t_e],Q) \leq \epsilon$.

b) $D(S[t_b{:}t_e],Q)$ is the minimum among all $D(S[t_b'{:}t_e'],Q)$, where $S[t_b'{:}t_e']$ is any subsequence that overlaps with $S[t_b{:}t_e]$.

\textit{Problem 3: top $k$ query.} %When $S$ is pre-recorded and its length is fixed, find the subsequence $S[t_b{:}t_e]$ of $S$ among all possible subsequences that has the smallest distance from the query sequence $Q$. That is $D(S[t_b{:}t_e],Q) \leq D(S[t_b'{:}t_e'],Q)$,  $\forall \ 0\leq t_b',t_e' \leq t$.
Find $k \geq 1$ disjoint subsequences of $S$ (each of them satisfies condition (b) in Problem 2) that have the smallest distances to the query sequence $Q$.

%\textbf{Problem 3: real-time monitoring}  When $S$ is a streaming data, and at every time tick a new value is appended to $S$. Report the subsequence $S[t_b{:}t_e]$ that satisfy condition a) in problem 2 in real-time. That is, report $S[t_b{:}t_e]$ at time tick $t_e$, once $D_{\text{norm}}(S[t_b{:}t_e],Q) \leq \epsilon$.
Due to the fact that all three problems are coupled to each other and considered to be general pattern matching problems, in this paper, we address them simultaneously.

\subsection{Limitations of traditional normalization approach}%es and fixed length sliding window}
An essential step preceding solving any of the three above mentioned problems is data normalization that allows pattern matching algorithms to reveal structural differences of considered signals. 
Performing z-normalization (that is the most common type of normalization) of a value $s_k$ is a scaling defined on a sequence $S[t_b{:}t_e] = \{s_{t_b},\dots,s_{t_e}\}$ that contains the value $s_k$ and is expressed as:
\begin{equation}
\label{eq:std_norm}
\begin{aligned}
s_{(t_b,t_e),k}' = \frac{s_k-\mu_{t_b,t_e}}{\sigma_{t_b,t_e}},\ \ k = t_b,\dots,t_e\\
\end{aligned}
\end{equation}
where $\mu_{t_b,t_e}$ and $\sigma_{t_b,t_e}$ are the mean and standard deviation of $S[t_b{:}t_e]$, respectively.
%Scaling by z-normalization ensures the transformation of the input signal to the one whose mean is approximately zero and variance is one.
% To perform z-normalization of a hidden in a stream subsequence it is required to scale each value using the mean and standard deviation obtained from the subsequence itself. We further call it as the intrinsic pattern interval since it corresponds to the intrinsic physical process interval.

\begin{figure}[htb]
\begin{minipage}[b]{1.0\linewidth}
  \centering
  \centerline{\epsfig{figure=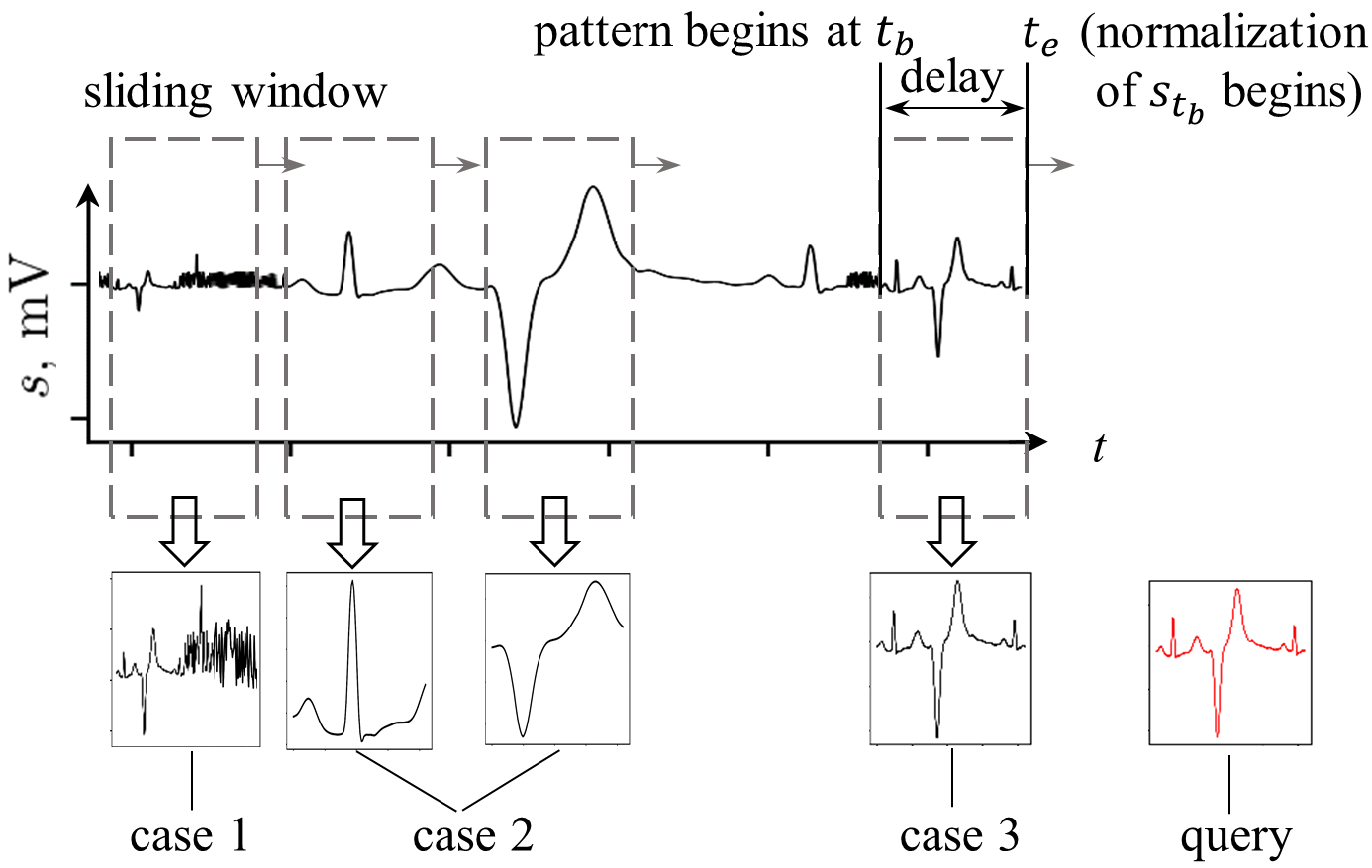,width=8.5 cm}}
  \caption{Scaling of ECG signal (taken from~\cite{DTW_million}). Cases 1 and 2 show scaling a subsequence using z-normalization with improper window length; case 3 shows a time delay due to the limitation of usual normalization approaches}
  \label{fig:sliding_window_errors}
\end{minipage}
\end{figure}

In most pattern matching scenarios the most challenging task is to define the time range $S[t_b{:}t_e]$ that is used for obtaining $\mu_{t_b,t_e}$ and $\sigma_{t_b,t_e}$~\cite{DTW_million}. 
Failure to select proper scaling subsequence might often result into significant distortions of structure of the signal being scaled.
The predominant practice to perform z-normalization approach is by using a fixed length sliding window. 
However, in scenarios where a pair of signals gets compared by DTW a too large or too short normalization window might result into redundant noise capturing or fragmentation of the matching subsequence, respectively. 
As a result, this leads to erroneous normalization and yields larger DTW dissimilarity between the query and the matching subsequence making the time-distorted subsequences undiscovered. 
Cases 1 and 2 in \Cref{fig:sliding_window_errors} demonstrate an improper normalization window that leads to erroneous scaled signals.
Note should be taken that finding an intrinsic pattern duration the normalization window in pattern matching tasks that employ DTW is practically ineffective as DTW assumes the possibility of time distortions between matching signals. 
The distortion level is usually unknown in advance in most of the real world applications. 

Additionally, even when the length of the sliding window coincides with the intrinsic length of the subsequence in the stream, the requirement of the entire window $S[t_b{:}t_e]$ in order to calculate scaling parameters introduces often unwanted time delays.
The proper normalization of the first value in the window $s_{t_b}$ can only be done when the last value $s_{t_e}$ is available, as shown in \Cref{fig:sliding_window_errors} (case 3). 
At time tick $t_e$ when $s_{t_e}$ is available, the existing methods either process $s_{t_b}$ only and process the subsequent $s_k (k>t_b)$ when $s_{t_e-t_b+k}$ ($t_e-t_b$ is the window length) is available~\cite{gong_si_fong_mohammed_2014} or try to process all of $\{s_{t_b},\dots,s_{t_e}\}$ at the time tick $t_e$~\cite{ISPRING,DTW_million}.
The former brings an additional time delay that is often undesired in many real-time scenarios and the latter increases the time complexity per time tick dramatically which can also cause a time delay in practice.

%\Cref{fig:sliding_window_errors}, \Cref{fig:prefix_procedure} and \Cref{fig:prefix_and_standard} all utilize a sequence taken from the ECG dataset in~\cite{DTW_million}.
%
%The naive scaling solution to account for the limitations of a fixed window z-normalization is to try all possible window length performing z-normalization based on resulting statistics. For a window with length $l_w$, the complexity for querying $Q$ on $S$ preforming sliding window z-normalization is $O(mtl_w)$. This makes the time complexity for problem 1 to be $O(mt^3)$ (or $O(mt^2)$ per time tick). This is infeasible for most applications. 
%
\section{Proposed pattern matching method}
\label{sec:proposed_method}

% NEEDS REWRITING
%To address the limitations of the traditional normalization that is based on the fixed length sliding window and 
%avoid subsequence discovery time delay introduced due to such type of normalization, 
In the following, we introduce a dynamic z-normalization method to address the limitations of the traditional normalization approaches %to find DTW distance between two subsequences
and propose a pattern matching method that utilizes this normalization concept to solve the three problems stated in \Cref{sec:problem_formulation}.

\subsection{Dynamic Time Warping (DTW)}
\label{ssec:DTW}
%DTW is the dissimilarity measure for time series and it has been shown that no other known distance measure reproducibly and significantly outperforms it~\cite{DTW_million}. 
The DTW distance between two sequences $X = \{x_0,\dots,x_{m-1}\}$ and $Y = \{y_0,\dots.y_{n-1}\}$ is defined as:
\begin{equation}
\label{eq:DTW_Distance}
\begin{aligned}
DTW(X,Y) =& D(m-1, n-1) \\
D(i,j) = ||x_i-y_j||& + \min \begin{cases}
D(i,j-1)\\
D(i-1,j)\\
D(i-1,j-1)\\
\end{cases}\\
D(-1,-1)=0,\  D(i,-1)&=D(-1,i) = +\infty\\
i=0,...,m-1;&\  j=0,...,n-1.\\
\end{aligned}
\end{equation}
and obtained by a bottom-up dynamic programming process, in which a time warping matrix of $m \times n$ %$m$ columns and $n$ rows 
is progressively constructed by appending a new column. Each cell $(i,j)$ of the matrix stores $D(i,j)$ - the minimum accumulative distance between subsequences $X[0{:}i]$ and $Y[0{:}j]$ which is achieved under the best alignment (an alignment is a set of contiguous matrix indices that defines a mapping between the elements of $X$ and $Y$). The matrix indices of the best alignment form a continuous warping path that starts in $(0,0)$ progressing to $(i,j)$ where the direction that the path heads to at every step depends on which one being the minimum in the $\min$ operation in \Cref{eq:DTW_Distance} of that step.
%The time complexity for DTW is $O(mn)$, while the space complexity is able to achieve $O(m)$ since only the latest two columns of the warping matrix are needed for computation.

\subsection{Dynamic z-normalization for DTW}
%CD: A hidden what?
Assume, that a pattern hidden in a stream has its intrinsic duration which might be different from the duration of the query signal.
To z-normalize such a signal preserving its structure a scaling of its every data point by~\Cref{eq:std_norm} is to be performed considering the parameters (the mean and the standard deviation) obtained on the pattern intrinsic duration interval.
In the following, we motivate our dynamic z-normalization approach that performs normalization of each data point of a hidden pattern $S[t_b{:}]$ in a streaming sequence (the pattern starts from $t_b$ and is arriving in real-time) without any assumptions or pre-knowledge of the intrinsic duration of the hidden pattern.

$\textbf{Idea 1: Prefix normalization}$ We define prefix normalization of a point $s_k$ on $S[t_b{:}t_e]$ as:
\begin{equation}
\label{eq:prefix_norm}
\begin{aligned}
s_{t_b,k}' =& \frac{s_k-\mu_{t_b,k}}{\sigma_{t_b,k}}, \quad k = t_b,\dots,t_e
\end{aligned}
\end{equation}
where $\mu_{t_b,k}$ is
\begin{equation}
\label{eq:prefix_mu}
\mu_{t_b,k} = \frac{1}{k+1-t_b}\sum_{l=t_b}^k s_l, \quad k=t_b,\dots,t_e
\end{equation}
and $\sigma_{k,t_b}$ is
\begin{equation}
\label{eq:prefix_sigma}
\sigma_{t_b,k} = \sqrt{\frac{1}{k+1-t_b}\sum_{l=t_b}^k s_l^2 - \mu_{t_b,k}^2}, \quad k=t_b,\dots,t_e
\end{equation}
The main idea of such a normalization is that for every data point $s_k$ to be normalized only its preceding signal is used in order to obtain its normalization parameters. 
%TODO: change the sentence below
%Therefore, $\forall j\geq k$, $s_k$ prefix-normalized in subsequence $S[i{:}j]$ will give the same value $s_{k,i}'$.
Therefore, this normalization is additive: %adding more values to the subsequence 
subsequent values do not affect previously normalized ones.
However, when $k$ is small, only few values are used to obtain $\sigma_{t_b,k}$ and $\mu_{t_b,k}$, and typically $\sigma_{t_b,k} < \sigma_{t_b,t_e}$ and $\mu_{t_b,k} \neq \mu_{t_b,t_e}$. 
As the result, such scaling leads to an amplification and an amplitude shift of the matching signal comparing to its scaled on its intrinsic pattern duration version. 
To compensate for that we introduce the amplification and the shift factors $\eta_k$ and $\delta_k$, correspondingly, that are defined as:
\begin{equation}
\label{eq:enlargement_factor}
\begin{aligned}
\eta_k = \frac{\sigma_{t_b,t_e}}{\sigma_{t_b,k}}, \quad \delta_k = \frac{\mu_{t_b,k}-\mu_{t_b,t_e}}{\sigma_{t_b,t_e}}, \quad
k = t_b,\dots,t_e.\\
\end{aligned}
\end{equation}
%being both large for small $k$. 
As $k$ increases, $\eta_k$ and $\delta_k$ converges to $1$ and $0$, respectively. 
By replacing the summations in \Cref{eq:prefix_mu} and \Cref{eq:prefix_sigma} with integration, the prefix normalization (\Cref{eq:prefix_norm}), as well as its amplification and shift factors (\Cref{eq:enlargement_factor}) can be also defined on continuous functions. 

$\textbf{Idea 2: Invariance properties}$ Let $f(x)$ be a continuous function defined on an interval $x \in [x_l,x_u]$. Let a continuous function $g(x)$ be defined by:
\begin{equation}
\label{eq:gx}
g(x) = C_2 f(C_1 (x+C_0))+C_3
\end{equation}
where $C_0$, $C_1$, $C_2$ and $C_3$ are the constants. The domain of definition for $g(x)$ is then $[x_l',x_u'] = [\frac{x_l}{C_1}-C_0,\frac{x_u}{C_1}-C_0]$.
Function $g(x)$ is similar to $f(x)$ in terms of shape (or structure), as it is transformed from $f(x)$ by stretching and shifting it both horizontally and vertically. 
The following invariants hold before and after the transformation:
\begin{equation}
\label{eq:invariants}
\eta_{x'}' = \eta_x,\quad \delta_{x'}'=\delta_x
\end{equation}
under the condition that $\frac{x'-x_l'}{x_u'- x_l'} = \frac{x-x_l}{x_u-x_l}$;
where $\eta_x$ and $\eta_{x'}'$ are the amplification factors for the prefix normalizations of $f(x)$ and $g(x)$ at $x$ and $x'$, respectively; $\delta_x$ and $\delta_{x'}'$ are the corresponding shift factors.
This derivation can be obtained by substituting \Cref{eq:gx} into \Cref{eq:enlargement_factor} (To maintain the flow of the analysis, we defer the detailed proof to \Cref{sec:proof_invariants}).

In case of discrete sequences, for $S[t_b{:}t_e]$ if it is similar to query sequence $Q$, the invariants in \Cref{eq:invariants} become:
\begin{equation}
\label{eq:invariants_discrete}
\eta_{k'}'=\eta_k,\quad \delta_{k'}'=\delta_k
\end{equation}
under the condition that $\frac{k'-t_b}{t_e-t_b} = \frac{k}{m-1}$; where $\eta_k$ and $\delta_k$ are the amplification and shift factors for the prefix normalization of $Q$, respectively and $\eta_k'$ and $\delta_k'$ are that of $S[t_b{:}t_e]$. It should be Noted that \Cref{eq:invariants,eq:invariants_discrete}
%\begin{equation}
%\eta_k = \eta_{k'}',\quad \delta_k = \delta_{k'}'
%\end{equation}

$\textbf{Idea 3: DTW embedding}$ The amplification and shift factors can be easily obtained for a query (template) sequence, while unknown for a pattern hidden in a stream since $\mu_{t_e,t_b}$ and $\sigma_{t_e,t_b}$ obtained from the pattern intrinsic duration interval are required. 
We propose to exploit the similarity between the query and the subsequence by \Cref{eq:invariants_discrete} and substitute the amplification and shift factors for the streaming sequence with that of the query sequence. 
To achieve that, the mapping between $k$ and $k'$ in \Cref{eq:invariants_discrete} is required. 
Therefore, we embed the normalization procedure into the DTW process, exploiting the sequence alignment of DTW to get the mapping. 
This procedure provides the normalized DTW distance ($D_{\text{norm}}$). 
We define $D_{\text{norm}}$ between the query $Q$ and the current sub-subsequence $S[t_b{:}t]$ of $S[t_b{:}t_e]$ as:
%\begin{equation}
%D_{\text{norm}}(S[t_b{:}t],Q) = \min_{0\leq k<m} D(t,k)
%\end{equation}
%where $D$ is defined as:
\begin{equation}
\label{eq:DTW_Distance_Normalized}
\begin{aligned}
D_{\text{norm}}(S[t_b{:}t],Q) =& D(t,m-1)\\
D(k',k) = d(k',k) & + \min \begin{cases}
D(k',k-1)\\
D(k'-1,k)\\
D(k'-1,k-1)\\
\end{cases}\\
D(-1,-1)=0,\  D(k',-1)&=D(-1,k) = +\infty\\
k'=t_b,...,t;& \quad k=0,...,m-1  
\end{aligned}
\end{equation}
where $d(k',k)$ is obtained by:
\begin{equation}
\label{eq:small_d}
d(k',k) = ||(\frac{s_{t_b,k'}'}{\eta_{k'}'} + \delta_{k'}') - (\frac{q_{0,k}'}{\eta_k} + \delta_k)||
\end{equation}
where $q_{0,k}'$ is the prefix normalized $q_k$ on $Q[0{:}k]$ and $s_{t_b,k'}'$ is the prefix normalized $s_{k'}$ on $S[t_b{:}k']$; $\eta_k$ and $\delta_k$ are the amplification and shift factors of $q_{0,k}'$ on the query $Q$; $\eta_{k'}'$ and $\delta_{k'}'$ are the amplification and shift factors of $s_{t_b,k'}'$ on the hidden pattern sequence $S[t_b{:}t_e]$.

In belief of similarity by \Cref{eq:invariants_discrete}, $\eta_{k'}'$ and $\delta_{k'}'$ are substituted by $\eta_k$ and $\delta_k$, so that $d(k',k)$ becomes:
\begin{equation}
\label{eq:d_t_k}
d(k',k) = ||\frac{s_{t_b,k'}'-q_{0,k}'}{\eta_k}||
\end{equation}
and the corresponding dynamically normalized value of $s_{k'}$ is:
\begin{equation}
\label{eq:dn_value}
s_{k'}'' = \frac{s_{t_b,k'}'}{\eta_k} + \delta_k
\end{equation}
Same as DTW distance, $D_{norm}$ is obtained by constructing a time warping matrix. When a new value $s_{t+1}$ arrives at time $t+1$, one new column is added to the matrix. 
According to the introduced dynamic z-normalization, we highlight its following features:
\begin{itemize}
\item The normalization process is additive that is aligned with DTW paradigm: 
consecutive value $s_{t+1}$ arriving at time $t+1$ does not affect previously scaled values.
%at the time tick $t+1$ the subsequence $S[i{:}t+1]$ generates a new column of 
%when the next value $s_{t+1}$ comes at time tick $t+1$, a new column is appended to the time warping matrix not affecting previously scaled values. 
In case of z-normalization with the fixed window (\Cref{eq:std_norm}) adding every new value to the window changes all previously normalized ones.
\item Each value $s_{k'}$ in the subsequence is dynamically z-normalized according to its mapped query value $q_k$.
\item Proper scaling of $s_{k'}$ is achieved when it is mapped to a truly similar $q_k$ so that $d(k',k)$ is small for each correct pair of $k'$ and $k$.
\item $D_{\text{norm}}(S[t_b{:}t],Q)$ reaches its minimum when $t$ increases to $t_e$ if $S[t_b{:}t_e]$ and $Q$ are truly similar.%, because the warping path can go through each right pair of mapping that give small distance.
\item As $S[t_b{:}t_e]$ gets more similar to $Q$, $D_{\text{norm}}(S[t_b{:}t],Q)$ becomes smaller since $d(k',k)$ gets smaller for each aligned pair of $k'$ and $d$.
\end{itemize}

\begin{figure}[htb]
\begin{minipage}[b]{1.0\linewidth}
  \centering
  \centerline{\epsfig{figure=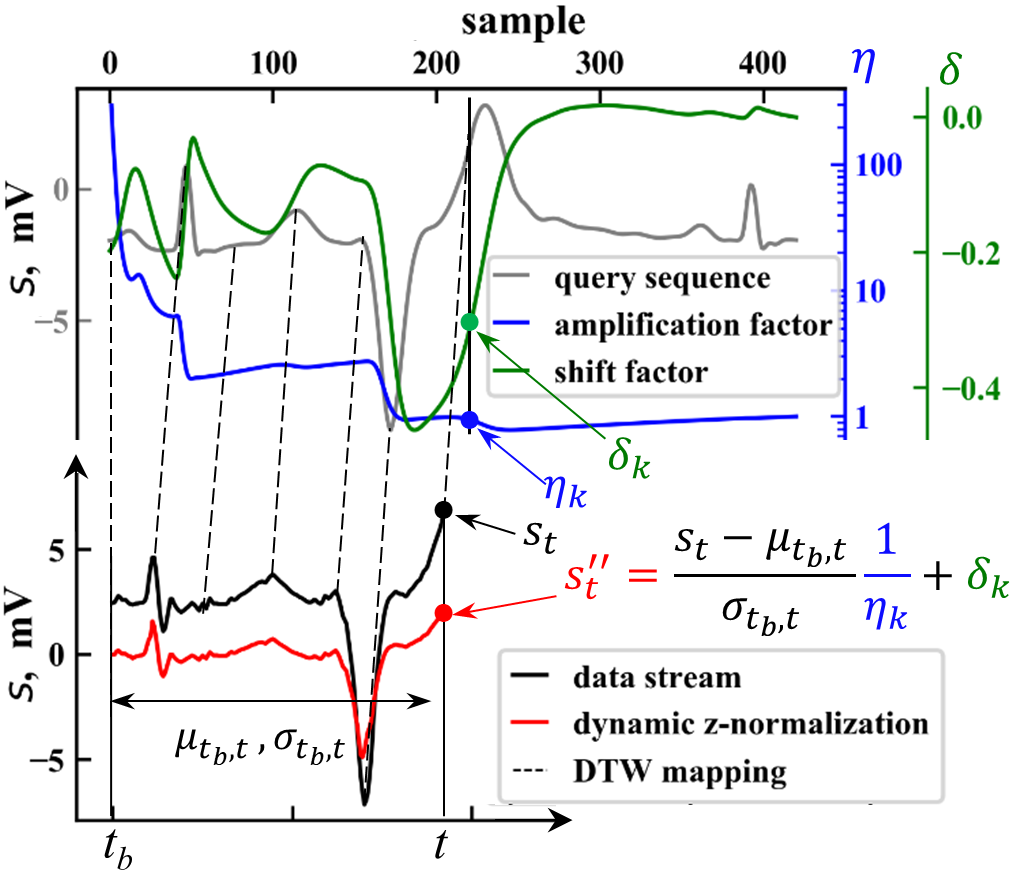,width=8.5 cm}}
%  \vspace{2.0cm}
  \caption{The proposed real-time dynamic z-normalization process on a ECG signal (from~\cite{DTW_million}). By integrating a normalization procedure into DTW mapping each value $s_t$ is dynamically z-normalized according to its mapped query value}
  \label{fig:prefix_procedure}
\end{minipage}
\end{figure}

\Cref{fig:prefix_procedure} illustrates the process of the proposed dynamic z-normalization when obtaining $D_{\text{norm}}(S[t_b{:}t_e],Q)$.

\begin{figure}[htb]
\begin{minipage}[b]{1.0\linewidth}
  \centering
 \centerline{\epsfig{figure=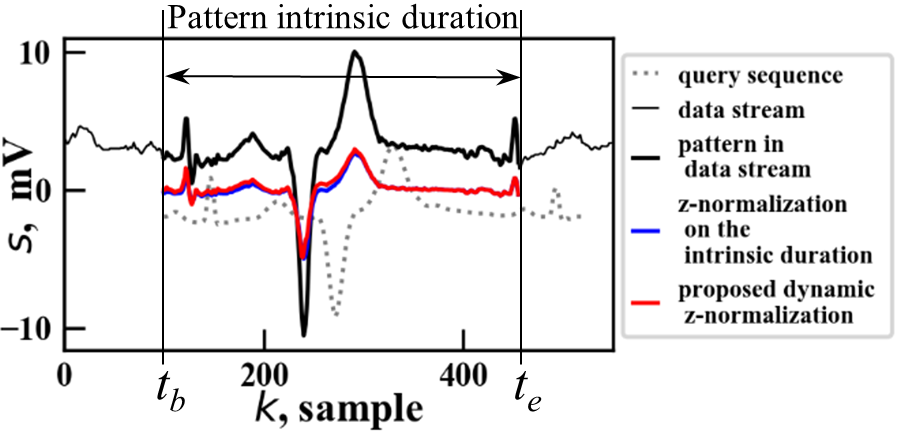,width=8.5 cm}}
  \caption{An example of two normalization approaches performed on a ECG signal sequence (taken from~\cite{DTW_million}): original signal is z-normalized by applying a window of the length of pattern intrinsic duration and by the proposed dynamic z-normalization. Since the hidden subsequence matches the query (template) sequence both normalization mechanisms provide similar results}
  \label{fig:prefix_and_standard}
\end{minipage}
\end{figure}

\Cref{fig:prefix_and_standard} demonstrates an example of the proposed dynamic z-normalization method on a ECG signal sequence. 

%\subsection{Uniform-scaling normalization for DTW}
%Dynamic normalization exploits the DTW mapping, while the slope of the warping path could also be informative. 
%If the warping path is more inclined to the data stream, that means on average every data point in the query sequence is mapped to multiple data points in the data stream, indicating a stretching in the data stream. 
%Similarly, if the warping path is more inclined to the query sequence, there would be a shrinking in the data stream. 
%We exploit this information to find the intrinsic window and normalize the data stream in that window.
%The only thing changes in \Cref{eq:DTW_Distance_Normalized} is:
%\begin{equation}
%d(l,k) = ||\frac{s_l - \mu_l}{\sigma_l} - \frac{q_k-\mu_Q}{\sigma_Q} ||
%\end{equation}
%where $\mu_Q$ and $\sigma_Q$ are the mean and standard deviation of the query sequence; $\mu_l$ and $\sigma_l$ are the mean and standard deviation of subsequence $S[i:e_l]$. $i + e_l$ is obtained by:
%\begin{equation}
%\begin{aligned}
%e_l =& \max(e_l', w_{min})\\
%e_l' =& \min(\frac{m(l-i)}{k}, w_{max})
%\end{aligned}
%\end{equation}
%where $w_{min}$ and $w_{max}$ are the minimum and maximum window length for normalization.

%The intuition behind is to try to use the slope of the warping path to predict the intrinsic window for normalization.

\subsection{Dynamic Normalization based Real-time Pattern Matching (DNRTPM) algorithm}
\label{ssec:DNRTPM}
In this section, we provide the details on our proposed Dynamic Normalization based Real-time Pattern Matching (DNRTPM) algorithm that is based on the introduced dynamic z-normalization principle and STWM concept~\cite{sakurai_faloutsos_yamamuro_2007}.
%The main advantage of DNRTPM with respect to methods that utilize STWM concept [LINKS to spring, nspring, ispring] is that the earlier employs 
%Due to the proposed dynamic normalization mechanism, the method is able to discover patterns of different scale in streaming data in real-time without any time delay and is not susceptible to time distortions.

\subsubsection{DNRTPM algorithm}
\begin{figure}[htb]
\begin{minipage}[b]{1.0\linewidth}
  \centering
  \centerline{\epsfig{figure=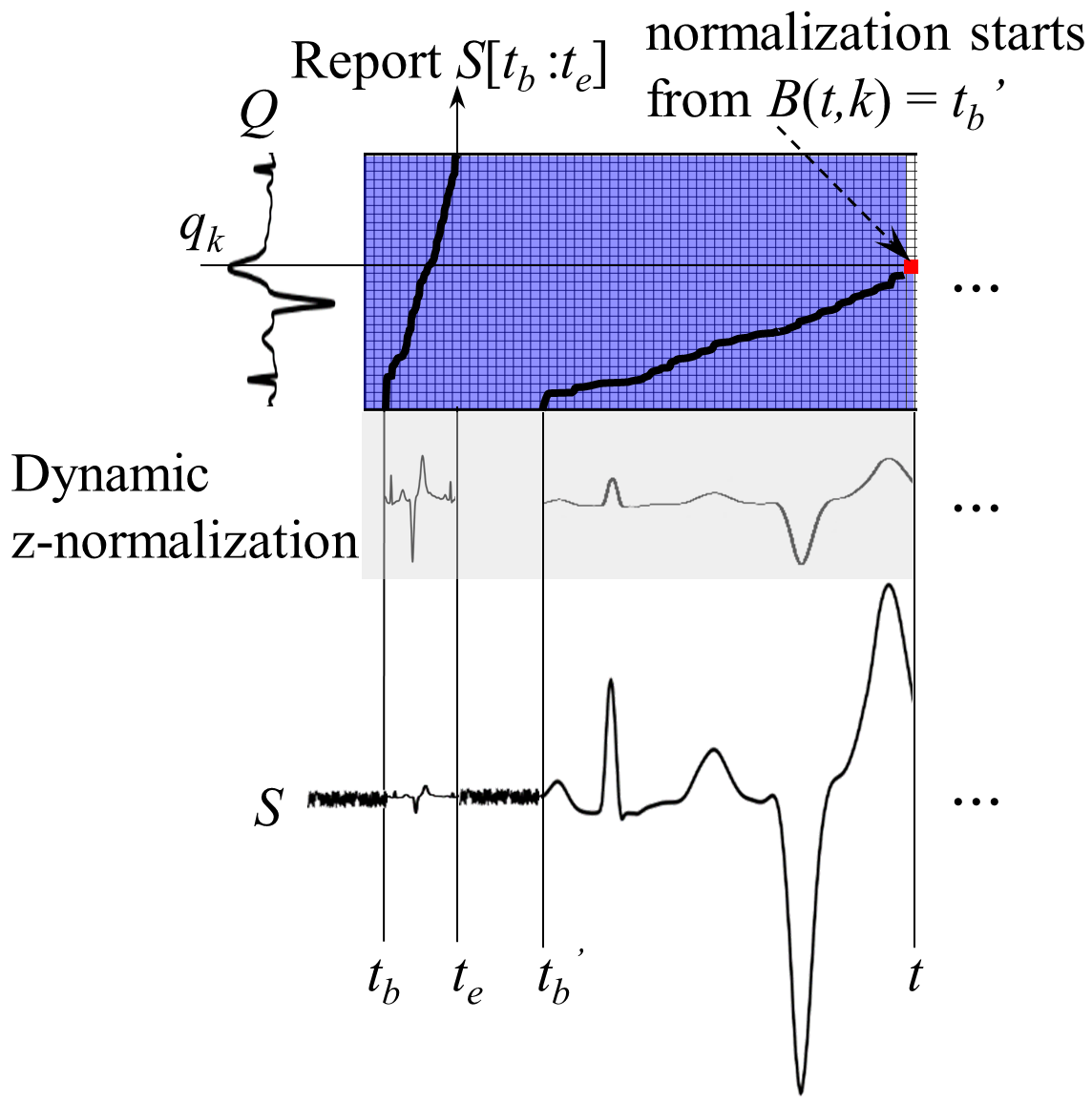,width=7 cm}}
%  \vspace{2.0cm}
  \caption{An example of subsequence matching procedure employed by DNRTPM}
  \label{fig:DNRTPM_matrix}
\end{minipage}
\end{figure}

The query (template) sequence $Q = \left\{q_0,q_1,\dots,q_{m-1}\right\}$ is first prefix-normalized by \Cref{eq:prefix_norm} to obtain the prefix-normalized query sequence $Q' = \left\{q_0',q_1',\dots ,q_{m-1}'\right\}$. The corresponding amplification factors $\eta = \{\eta_0,\eta_1, \dots,\eta_{m-1} \}$ are obtained by \Cref{eq:enlargement_factor}.
%As for the streaming sequence $S$, the value $s_t$ at the latest time tick is prefix-normalized on the fly when it is being used.
The main operational principle of DNRTPM is illustrated in \Cref{fig:DNRTPM_matrix}. 
For the given query and the incoming data stream DNRTPM constructs a STWM (colored in blue) in order to align the query with the streaming subsequence and return their distance.
Each cell in the STWM stores two values: $B(t,k)$ and $D(t,k)$.
$B(t,k)$ denotes the index of the beginning point of the possible candidate subsequence with a warping path (colored in black in \Cref{fig:DNRTPM_matrix}) that goes through cell $(t,k)$.
$D(t,k)$ is the accumulative distance of the warping path and is obtained by:
\begin{equation}
\label{eq:spring_D}
\begin{aligned}
D(t,k) =& \min \begin{cases}
D'(t,k-1)\\
D'(t-1,k)\\
D'(t-1,k-1)\\
\end{cases}\\
D'(i,j) =& ||\frac{s_{B(i,j),t}'-q_{0,k}'}{\eta_k}|| + D(i,j)\\
i = t-1,&t;\  j=k-1,k.\\
D(t,-1)& = 0,\  D(-1,k) = +\infty.\\
t = 0,&\dots,n;\  k = 0,\dots,m-1.
\end{aligned}
\end{equation}
%\begin{equation}
%\label{eq:spring_D_prime}
%\begin{aligned}
%D'(i,j) =& \min_{0\leq q \leq m-1}(||\frac{s_t -M(i,j)}%{SD(i,j)} - T_{qk}'||) + D(i,j)\\
%\end{aligned}
%\end{equation}
where $q_{0,k}'$ is the $k$th prefix-normalized value of the query sequence; 
$s_{B(i,j),t}'$ is $s_t$ prefix-normalized on subsequence $S[B(i,j){:}t]$ by \Cref{eq:prefix_norm}. 
%$\eta_k$ is the amplification factor.
% of the query sequence obtained by \Cref{eq:enlargement_factor} and used here to counteract the amplification brought by prefix-normalization; 
%$D(t,-1) = 0$, $D(-1,k) = +\infty$. 
%The cells that have already been, are being and will be processed next are colored in blue, red and white colors, respectively. 
When a new value arrives at the time tick $t+1$ a new column is appended to the right side of STWM. $B(t,k)$ is obtained as follows:
\begin{equation}
\label{eq:spring_B}
\begin{aligned}
B(t,k) =& \begin{cases}
B(t-1,k),& \text{if}\ D(t,k) = D'(t-1,k)\\
B(t,k-1),& \text{if}\ D(t,k) = D'(t,k-1)\\
B(t-1,k-1),& \text{if}\ D(t,k) = D'(t-1,k-1)\\
\end{cases}\\
B(t,0) & = t;\ t = 0,\dots,n;\  k = 0,\dots,m-1.\\
\end{aligned}
\end{equation}
%where $B(t,0) = t$.
In \Cref{eq:DTW_Distance_Normalized}, the result of the min operation does not affect the value of $d(k',k)$ since the beginning index is fixed as $t_b$(\Cref{eq:small_d}), while in \Cref{eq:spring_D} the beginning index $B$ is different for the three options on the right side of the min operation.
Therefore, in \Cref{eq:spring_D} this difference is considered before taking the minimum.

In order to efficiently calculate $s_{B(i,j),t}'$ a list of prefix summations $\text{PS}$ and a list of prefix summations of squares $\text{PSS}$ are maintained. At the current time tick $t$, $\text{PS}$ and $\text{PSS}$ are defined as:
\begin{equation}
\begin{aligned}
\text{PS} =& \left\{\text{ps}_{(B(t-1,min)-1)},\dots,\text{ps}_t\right\}\\
\text{PSS} =& \left\{\text{pss}_{(B(t-1,min)-1)},\dots,\text{pss}_t\right\}\\
B(t-1,min) =& \min_{0 \leq i < m} B(t-1,i)\\
\end{aligned}
\end{equation}
where $\text{ps}_i$ and $\text{pss}_i$ are the prefix summation and the prefix sum of squares till time tick $i$. 
Values $\text{ps}_i$ and $\text{pss}_i$ are obtained by:
\begin{equation}
\label{eq:prefix_sums}
\begin{aligned}
\text{ps}_i =& \text{ps}_{i-1} + s_i \\
\text{pss}_i =& \text{pss}_{i-1} + s_i^2\\
i =& 0,\dots,t; \  \text{ps}_{-1} = \text{pss}_{-1} = 0.\\
\end{aligned}
\end{equation}
$\text{PS}$ and $\text{PSS}$ allow to efficiently obtain $s_{B(i,j),t}'$ by:
\begin{equation}
\label{eq:normalize_st}
\begin{aligned}
s_{B(i,j),t}' =& \frac{s_t - \mu_{B(i,j),t}}{\sigma_{B(i,j),t}} \\
\mu_{B(i,j),t} =& \frac{\text{ps}_t - \text{ps}_{(B(i,j)-1)}}{t-B(i,j)+1}\\
\sigma_{B(i,j),t} =& \sqrt{\frac{\text{pss}_t-\text{pss}_{(B(i,j)-1)}}{t-B(i,j)+1} - \mu_{B(i,j)}^2,t}.\\ 
\end{aligned}
\end{equation}
When the consecutive value $s_{t+1}$ arrives at time tick $t+1$, $\text{PS}$ is updated by removing $\left\{\text{ps}_{(B(t-1,min)-1)},\dots,\text{ps}_{(B(t,min)-2)}\right\}$ from its beginning and appending $\text{ps}_{t+1}$ to its end. $\text{PSS}$ is updated in the same way as  $\text{PS}$. $\text{PS}$ and $\text{PSS}$ are implemented by circular buffer deque to achieve constant $O(1)$ time appending and removing at the beginning or end, and also constant $O(1)$ random access in \Cref{eq:normalize_st}. The size of the circular buffer can be changed dynamically whenever it is required. This technique to incrementally obtain the mean and standard deviation resembles the one used in online z-normalization~\cite{DTW_million}, but the latter only allows to get the mean and standard deviation of the preceding subsequence with a predefined and fixed length, while ours supports flexible length.

In situations that require real-time responding, at the current time tick $t$, if $D(t,m-1)$ is smaller than a predefined threshold $\epsilon$, the subsequence $S[B(t,m-1){:}t]$ is reported immediately. 
However, it is possible that there are multiple overlapping subsequences that all have a distance smaller than $\epsilon$. 
In case of disjoint query task (that reports non-overlapping subsequences) %is wanted in situations where real-time responding is less important than accuracy. 
the subsequence $S[B(t,m-1){:}t]$ is only reported after confirming that all the upcoming subsequences that overlap with itself provide larger distance. 
This is achieved by keeping current minimum distance $D_{min}$ and the corresponding subsequence $S_{opt}$ (starts at $t_s$ and ends at $t_e$)  reporting $S[B(t,m-1){:}t]$ only when:
\begin{equation}
\label{eq:disjoint_query}
\forall_k, D(t,k) \geq D_{min} \lor B(t,k) > t_e.
\end{equation} 
$D_{min}$ and $S_{opt}$ are updated as the current distance $D(t,m-1)$ and subsequence $S[B(t,m-1){:}t]$ when $D(t,m-1) < D_{min}$.

The pseudo-code of the proposed DNRTPM algorithm for disjoint query problem is summarized in~\Cref{alg:DNRTPM} in \Cref{sec:pseudocode}. 
The adoption to real-time monitoring problem is done by reporting ($D(t,m-1)$, $B(t,m-1)$, $t$) as soon as condition in \Cref{alg_line:real_cond} of~\Cref{alg:DNRTPM} is satisfied. 
Furthermore, the adoption for top $k$ query problem is done by keeping the best $k$ discovered subsequences reported from \Cref{alg_line:output} of~\Cref{alg:DNRTPM} and maintaining $\epsilon$ as the smallest distance for the $k$ subsequences.

\subsubsection{Space and time complexity}
According to \Cref{eq:spring_D} and \Cref{eq:spring_B}, as well as to \Cref{alg:DNRTPM}, only the values $D$ and $B$ corresponding to the columns of STWM of the current and previous time tick are needed to be kept in memory. 
Additionally, two deques $\text{PS}$ and $\text{PSS}$ whose length is comparable to $m$ are required. 
As the result, the space complexity of the proposed method is $O(m)$.

The time complexity to fill each cell of the warping matrix is $O(1)$ leading to $O(m)$ per time tick or $O(mt)$ for the whole process.

\section{Experiments}
\label{sec:experiments}
In this section, we evaluate our proposed DNRTPM algorithm and compare its performance with the other state-of-the-art pattern discovery methods, namely, NSPRING, ISPRING, UCR-US and UCR-DTW applying them on several synthetic and real-world datasets. 

For that, at first, %we demonstrate the importance of normalization procedure for time series pattern matching problems and 
we show the critical influence of window length for z-normalization as well as reveal that proposed dynamic z-normalization is nearly identical to z-normalization that is performed on the intrinsic pattern duration window length.
Further, we demonstrate the robustness of DNRTPM to uniform scaling on time axis.
By applying distortions to data of real-world UCR archive ~\cite{UCRArchive2018} we simulate realistic practical pattern matching scenarios evaluating operational performance of the proposed and the state-of-the-art methods.
After that, we evaluate the performance of all the methods on real-world mouse dynamics data.
Finally, we test the scalability and time delay of all methods on ECG data.
The evaluation in all experiments is mainly done by solving a top $k$ query problem avoiding setting a predefined threshold that is required for real-time monitoring and disjoint query tasks and strongly depends on the dataset and the domain it is coming from.
Practically, the threshold can be tuned in order to make real-time monitoring or disjoint query tasks reporting only $k$ non-overlapping subsequences providing similar result as top $k$ query task provides.

All experiments are performed using a PC with Intel Xeon(R) Gold 5120 CPU 2.20GHz $\times$ 28 with 16GB 2666MHz $\times$ 6 RAM. 

\subsection{Experiment 1: Dynamic z-normalization}
\label{ssec:DN_VS_ZN}
In this experiment, we compare two types of normalization for DTW: the proposed dynamic z-normalization and the traditional z-normalization with fixed length window. 

\begin{figure*}[htb]
%\begin{minipage}[b]{1.0\linewidth}
  \centering
  \centerline{\epsfig{figure=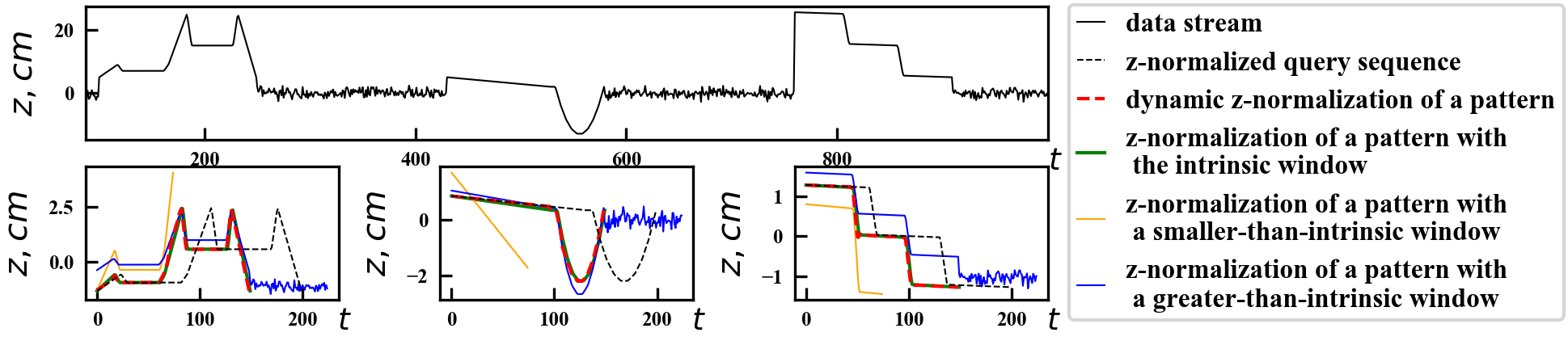,width=\textwidth}}
%  \vspace{2.0cm}
  \caption{Proposed dynamic z-normalization vs sliding window z-normalization with different window lengths. An improper window length makes z-normalization unable to bring the query and the hidden in the stream pattern to the same scale resulting in higher DTW distance. Dynamic z-normalization and z-normalization with the intrinsic pattern duration window are nearly identical and both are at the same scale in z direction as the normalized query sequence}
    \label{fig:DN_VS_ZN}
%\end{minipage}
\end{figure*}

For that we create three geometric shapes: cat, spoon and stairs that are used as query sequences. 
The data stream is created by concatenating sequences of white noise with a duration of $180$ samples and the three shapes after adding distortion by scaling in both z (amplitude) direction by $200\%$ and t direction by $75\%$ and shifting in z direction by $5$ as shown in \Cref{fig:DN_VS_ZN}.
To demonstrate the effect of window size on the result of z-normalization we run sliding window z-normalization on the resulting data stream with different window sizes: the intrinsic pattern duration window, a $50\%$ greater window and a $50\%$ smaller window. We run the proposed DNRTPM with best query (top $1$ query) setting on the resulting data stream and report normalized pattern subsequences in \Cref{fig:DN_VS_ZN} by monitoring subsequence's values applying \Cref{eq:dn_value}.
%The dynamic normalization mechanism of DNRTPM is embedded into its distance calculation process (\Cref{eq:small_d}) and is adapted dynamically based on the DTW mapping. 
% We keep track of how each value is normalized by \Cref{eq:dn_value} and consider the dynamically normalized values used for calculating the accumulative distance of the reported subsequence as the result of dynamic z-normalization.
In cases when a subsequence data point $s_i$ is mapped to multiple data points of the query sequence contributing multiple times to the accumulative distance % and each time it could be normalized differently.
we consider the average of the multiple normalization of $s_i$ to represent its dynamically normalized value.
The normalized pattern subsequences in data stream by z-normalization with different window sizes and by dynamic z-normalization are shown in \Cref{fig:DN_VS_ZN} while the DTW distances between them and their corresponding z-normalized original shapes are provided in \Cref{table:DN_vs_ZN}.

\begin{table}[!htb]
\begin{tabular}{|l|l|l|l|l|}
\hline
\multicolumn{2}{|l|}{} & cat & spoon & stairs \\ \hline
\multirow{3}{*}{\begin{tabular}[c]{@{}l@{}}z-norm-\\ alization\end{tabular}} & \begin{tabular}[c]{@{}l@{}}smaller-than-intrinsic\\  window\end{tabular}  & 77.25 & 36.57 & 36.96 \\ \cline{2-5} 
 & \begin{tabular}[c]{@{}l@{}}greater-than-intrinsic\\  window\end{tabular}& 84.91 & 24.22 & 30.26 \\ \cline{2-5} 
 & intrinsic window & 3.96 & \textbf{1.84} & \textbf{0.67} \\ \hline
\multicolumn{2}{|l|}{dynamic z-normalization} & \textbf{3.70} & 2.28 & 1.09 \\ \hline
\end{tabular}
\caption{DTW distances between the z-normalized original shapes and their corresponding distorted subsequences in data stream normalized by proposed dynamic z-normalization and traditional z-normalization}
\label{table:DN_vs_ZN}
\end{table}

An important observation regarding the importance of proper time series normalization comes from \Cref{fig:DN_VS_ZN}: 
improper normalization window length makes it miss parts of the hidden pattern or include unwanted noise signal. 
Both cases lead to a degradation of the performance of consequent pattern matching approach.
In case of z-normalization, the non-intrinsic pattern duration window does not allow to bring the signal to the right scale damaging the performance of the following pattern matching method.
As the result, sliding window based methods (e.g. z-normalization) require a proper window length to be set that can be practically impossible as the distortion level on time axis is unknown and can vary over time.
As demonstrated in \Cref{fig:DN_VS_ZN} and \Cref{table:DN_vs_ZN}, the proposed dynamic z-normalization is nearly identical to z-normalization applied on the window that equals to the length of the pattern.

\vspace{-0.25cm}
\subsection{Experiment 2: Pattern matching with synthetic shapes}
\label{ssec:exp_shapes}
In this experiment, we utilize the shapes generated in the previous experiment and their upside-down (reflected across time axis) versions to demonstrate DNRTPM's robustness to uniform scaling as well as to compare its performance to that of NSPRING, ISPRING, UCR-US and UCR-DTW.

%Uniform scaling factor $\lambda$ is defined as the length of the query sequence divided by the intrinsic length of the pattern in data stream.

We use z-normalized original shapes as query sequences.
Each shape is then uniformly scaled (by applying linear interpolation and re-sampling) in t direction to have the length of $\frac{m}{\lambda}$, where $m$ is the length of the original shape and $\lambda$ is a varying uniform scaling factor.
The data stream is then created by concatenating $30$ uniform-scaled shapes of each type with white noise of length $\frac{2m}{\lambda}$ between every two shapes resulting into $180$ shapes hidden in the data stream.
Every shape in the data stream is then randomly scaled in z direction by an amplification factor drawn from uniform distribution U(0,10) and randomly shifted in z direction by a factor drawn from U(-5,5).

We perform top 30 disjoint query task on the resulting data stream using each of the six z-normalized original shapes as the query sequences.
When querying a particular shape on the data stream a corresponding pattern hidden in the data stream is considered to be retrieved (or recalled) when the overlapping percentage between the query and any of the $30$ reported subsequences is greater than a predefined overlapping percentage ($\alpha$) that is defined between two subsequences $S[i{:}j]$ and $S[i'{:}j']$ according to~\cite{niennattrakul_wanichsan_ratanamahatana_2010} as:
\vspace{-0.1cm}
\begin{equation}
\label{eq:overlapping_percentage}
\begin{aligned}
\alpha = \begin{cases}
\frac{\min(j,j') - \max(i,i')+1}{\max(j,j')-\min(i,i')+1},&\text{if} \min(j,j') \geq \max(i,i')\\
0, &\text{if}\min(j,j') < \max(i,i')\\
\end{cases}
\end{aligned}
\end{equation}

\vspace{-0.1cm}

\begin{figure}[!htb]
\begin{minipage}[b]{1.0\linewidth}
  \centering
  \centerline{\epsfig{figure=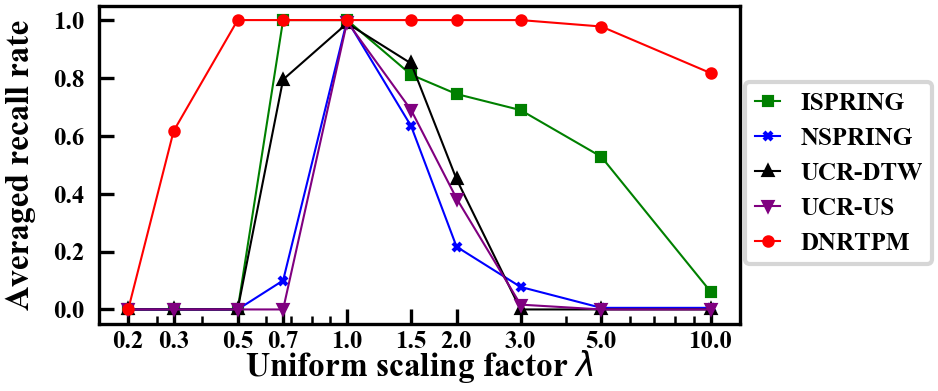, width=8.5cm}}
%  \vspace{2.0cm}
  \caption{Recall rate at different uniform scaling factors averaged over six top 30 queries for the six shapes}
  \label{fig:shapes_distortion}
\end{minipage}
\end{figure}

Recall rate is defined as the percentage of hidden in the data stream patterns corresponding to the query that are retrieved.
In case of a very low threshold $\alpha$, a naive method that reports the whole stream results into $100\%$ recall. 
We set $\alpha=50\%$ (in all relevant experiments) allowing UCR-DTW to report valid matches since it is able to only report fixed length patterns.
%However, a high threshold $\alpha$ does not allow UCR-DTW to report many valid matches (since it only reports fixed length subsequences). 
%As a trade-off, we set the threshold to be $50\%$ for all relevant experiments in this paper.
For every shape, there are 30 of its corresponding hidden patterns, so the recall rate, precision (percentage of correctly retrieved patterns) and F1 score (harmonic mean of precision and recall) in every top 30 query are equal.
UCR-US requires a predefined maximum scaling factor that is unknown in practice, so we set the maximum scaling factor in UCR-US as 200\% in all relevant experiments. 
For all the considered pattern matching methods we do not set a any global constraints to avoid additional parameters as well as the risk of negative effect to the performance due to the existence of the uniform scaling.
%Given the existence of uniform scaling in time axis, applying global constrain will make the methods unable to match a subsequence whose length is considerably different from the query, so no global constrain is used for any considered pattern matching method.
We perform the experiment for various uniform scaling factors ranging it from $0.25$ to $10$.
The recall rate averaged over all shapes for all methods for different uniform scaling factors is shown in \Cref{fig:shapes_distortion}.

The values of the recall rate on \Cref{fig:shapes_distortion} demonstrate that due to the window-free nature of dynamic z-normalization principle, the proposed DNRTPM is robust to both distortions in t axis (uniform scaling in time axis) and z axis. 
According to \Cref{fig:shapes_distortion}, when querying shapes with distortions in t and z axises, DNRTPM consistently provides recall rate of $1$ even under very large amount of uniform scaling ($\lambda$ ranges from $0.5$ to $3$)
The recall rate for other state-of-the-art methods decreases rapidly with the presence of uniform scaling ($\lambda \neq 1$) mostly due to their normalization mechanism with an improper window length that is unable to bring the hidden subsequences to the right scale. %(additionally, UCR-DTW only reports subsequences whose length equals to the one of the query sequence). 
%Note should be taken that ISPRING performs better than NSPRING and UCR-DTW since the datastream does not contain any extreme values making min-max normalization (that ISPRING is employing) comparable to z-normalization. 
For the low uniform scaling factor values ($\lambda < 0.5$), the performance of all methods  decreases. 
According to \Cref{eq:spring_D} for DNRTPM the distance between the query $Q$ of length $m$ and any subsequence $S[i{:}j]$ is a summation of $n_d$ distances each obtained by \Cref{eq:small_d}, where $n_d >= \max(j-i+1,m)$. 
When $\lambda$ is much smaller than $1$, for a pattern subsequence $S[i'{:}j']$, $j'-i'$ is much greater than $m$ and $n_d >= j'-i'$. 
In this case, DNRTPM favors shorter subsequences, since this way $n_d$ can be smaller and there is a smaller number of distances obtained by \Cref{eq:small_d} in the summation, so that the distance is smaller. 
This causes DNRTPM to find relatively shorter sequences than the true longer patterns in data stream when $\lambda << 1$. 
Similarly to that ISPRING and NSPRING favor shorter sequences when $\lambda < 1$ which when adding the effect of improper normalization makes their performance decrease more rapidly than that of $\lambda > 1$. Surprisingly, although UCR-US is designed to support uniform scaling, it doesn't perform well. This is because it also suffers from the favoritism of shorter sequences and it requires a predefined maximum scaling factor which is unknown in practice and the Euclidean distance it utilizes is not robust to the possible non-linear local distortion caused by the re-scaling inside UCR-US.

\subsection{Experiment 3: Pattern matching on UCR archive}
\label{ssec:Test_UCR}
In this section, we compare DNRTPM, NSPRING, ISPRING, UCR-US and UCR-DTW by performing top k query task on UCR archive~\cite{UCRArchive2018}, which contains 128 real datasets from various fields and is frequently used for assessing the performance of pattern matching algorithms. 
The data in UCR archive are all well-prepared: every dataset consists of pre-segmented short sequences for which the class labels are provided.
Besides that all sequences are z-normalized and within the same dataset exhibit equal length (hold for nearly all datasets).
Long data streams were segmented into short sequences by hand~\cite{hu2013time, koch2010gesture}
%Short sequences are commonly segmented from long data streams by hand~\cite{hu2013time, koch2010gesture} 
and the recording processes were contrived to make each pattern subsequence having roughly the same length~\cite{hu2013time, PAMAP, CMU_database, ratanamahatana2004making, yang2009distributed}. 
Unfortunately, in most real-life scenarios, especially in real-time pattern matching tasks, such thorough data preparation is unfeasible or can only be done with significant effort.

In order to simulate realistic streaming scenarios, we performed the following adjustments on every UCR dataset: Every sequence $S$ in a dataset is randomly scaled in z direction by an amplification factor drawn from U(0,10) and then randomly shifted in z direction by a factor drawn from U(-5,5).
% \textbf{Step 2.} Subsequently, foer very $S$, two random sequences (used for distraction) drawn from another random dataset in UCR archive are uniform-scaled in t direction to the same length as $S$, then attached to the beginning and end of $S$, respectively; 
Subsequently, every sequence is uniformly-scaled in t direction by a random rate $\lambda$ (or $\frac{1}{\lambda}$ by a chance of $50\%$) drawn from U(1,2). 
Finally, all sequences in the training subset are concatenated into a single long data stream by a random order, while the sequences in the testing set are used as query sequences.

For every query sequence $Q$ in every dataset, we perform a top $k$ disjoint query on the data stream created from the same dataset, where $k$ is the number of sequences in the training set that belong to the the same class as $Q$.

Despite it has been shown that UCR-DTW is able to perform top 1 query on datasets with a trillion datapoints in 34 hours~\cite{DTW_million}, in our experiments, UCR-DTW is slow on some datasets in our experiments. The reason for that is the difference in the experiment setting: 1) a tight global constraint of $5\%$ that was applied on the warping path in~\cite{DTW_million} while in our current experiment we apply no warping path constraint to avoid additional parameters; 2) top 1 query was performed in~\cite{DTW_million} while, here, we have adopted top k query (where k can be as large as 300), which limits the power of pruning lowerbounds (as in~\cite{DTW_million}); 3) the above mentioned adjustment increases the equivalent size of the datasets. For example, for the UWaveGestureLibraryAll dataset, the time complexity is equivalent to a single top k search on a sequence of 3032,951,040 data points. Due to this fact, we sort all datasets by UCR-DTW's time complexity, that is: number of testing sequences $\times$ square of the length of the testing sequence $\times$ number of the training sequences $\times$ length of the training sequences, then pick the smallest $80$ datasets that provide the lowest execution time.

To evaluate the performance of pattern matching approaches we use the following metrics: recall rate, %overlapping percentage (AoD),
 DTW distance (between the z-normalized query and the z-normalized retrieved pattern) and query time.% and reporting delay. 
 Recall is the percentage of the pattern subsequences in the data stream with the same class label as $Q$ that are found by top $k$ query. Similar to \Cref{ssec:exp_shapes} recall, precision and F1 are equal. 
%AoD is the average overlapping percentage $\alpha$ of the found sequences, which measures the quality of the found sequences~\cite{niennattrakul_wanichsan_ratanamahatana_2010}.
DTW distance shows dissimilarity under DTW between a z-normalized reported subsequence and the z-normalized query so it measures the quality of the retrieved subsequences. 
Query time is the wall clock time of a single query. 
%Reporting delay measures the difference between the time when the whole hidden pattern gets available to query and its reporting time. 
%To measure the reporting delay, firstly, top $k$ query task is performed to get the $k$th smallest distance and then this distance is set as the threshold for a stream monitoring task. 
%Note should be taken that the disjoint query by \Cref{eq:disjoint_query} is not used here and discovered subsequence is reported as soon as it has smaller than the threshold distance. 
%For a pattern hidden in the data stream, there can be multiple reported subsequences that overlap with it. 
%The reporting delay is the difference between the ending time tick of the pattern and the reporting time tick of the first subsequence that overlaps with the pattern (both time ticks are in the data stream clock) divided by the length of the pattern.
As in \Cref{ssec:exp_shapes}, no global constraints are used for any considered method. The performance measures for NSPRING, ISPRING, UCR-US, UCR-DTW and the proposed DNRTPM averaged over all datasets are shown in \Cref{table:ucr_summary}.

\Cref{table:ucr_summary} demonstrates that the proposed DNRTPM is able to achieve higher subsequence discovery performance than the other methods. 
In particular, DNRTPM achieves significantly higher recall as well as provides quality retrieval of subsequences according to DTW distance metric.
%\begin{figure}[!htb]
%\begin{minipage}[b]{1.0\linewidth}
%  \centering
%  \centerline{\epsfig{figure=recall_ucr.png,width=8.5 cm}}
%%  \vspace{2.0cm}
%  \caption{Recall on datasets in UCR archive.}
%  \label{fig:ucr_recall}
%\end{minipage}
%\end{figure}
%
%\begin{figure}[!htb]
%\begin{minipage}[b]{1.0\linewidth}
%  \centering
%  \centerline{\epsfig{figure=AoD_ucr.png,width=8.5 cm}}
%%  \vspace{2.0cm}
%  \caption{AoD on datasets in UCR archive.}
%  \label{fig:ucr_aod}
%\end{minipage}
%\end{figure}
%
%
%\begin{figure}[!htb]
%\begin{minipage}[b]{1.0\linewidth}
%  \centering
%  \centerline{\epsfig{figure=DTW_distance_ucr.png,width=8.5 cm}}
%%  \vspace{2.0cm}
%  \caption{Normalized DTW distance on datasets in UCR archive.}
%  \label{fig:ucr_dtw_norm}
%\end{minipage}
%\end{figure}
%
%
%\begin{figure}[!htb]
%\begin{minipage}[b]{1.0\linewidth}
%  \centering
%  \centerline{\epsfig{figure=Query_time_ucr.png,width=8.5 cm}}
%%  \vspace{2.0cm}
%  \caption{Query time on datasets in UCR archive.}
%  \label{fig:ucr_query_time}
%\end{minipage}
%\end{figure}
%
%\begin{figure}[!htb]
%\begin{minipage}[b]{1.0\linewidth}
%  \centering
%  \centerline{\epsfig{figure=Reporting_delay_ucr.png,width=8.5 cm}}
%%  \vspace{2.0cm}
%  \caption{Reporting delay on datasets in UCR archive.}
%  \label{fig:ucr_delay}
%\end{minipage}
%\end{figure}

\begin{table*}[htb]
\begin{tabular}{|l|l|l|l|l|l|l|}
\hline
\multicolumn{2}{|l|}{} & ISPRING & NSPRING & UCR-DTW & UCR-US& DNRTPM \\ \hline
\multirow{2}{*}{Recall} & averaged value& 0.34 & 0.36 & 0.29 &0.27 & \textbf{0.42} \\ \cline{2-7} 
 & best on x\% datasets& 8\% & 14\% & 13\% &7\% & \textbf{58\%} \\ \hline
%\multirow{2}{*}{AoD} & \begin{tabular}[c]{@{}l@{}}averaged\\  value\end{tabular} & 0.79 & 0.79 & 0.77 &\textbf{0.84}& \textbf{0.84} \\ \cline{2-7} 
% & \begin{tabular}[c]{@{}l@{}}best on x\% \\ datasets\end{tabular} & 5\% & 11\% & 12\% &\textbf{37\%}& 35\% \\ \hline
\multirow{2}{*}{DTW distance} &averaged value& 24.84 & 26.36 & 30.88 & 25.49&\textbf{21.05} \\ \cline{2-7} 
 & best on x\% datasets & 21\% & 16\% & 6\% & 8\%&\textbf{49\%} \\ \hline
%\begin{tabular}[c]{@{}l@{}}Reporting\\  delay\end{tabular} & \begin{tabular}[c]{@{}l@{}}averaged \\ value\end{tabular} & \textbf{\begin{tabular}[c]{@{}l@{}}no \\ delay\end{tabular}} & 0.73 & \textbf{\begin{tabular}[c]{@{}l@{}}no \\ delay\end{tabular}} & \textbf{\begin{tabular}[c]{@{}l@{}}no \\ delay\end{tabular}} &\textbf{\begin{tabular}[c]{@{}l@{}}no \\ delay\end{tabular}} \\ \hline
Query time& averaged value (seconds) & 5.48 & \textbf{0.41} & 44.04 & 4.18 &0.92 \\ \hline
\end{tabular}
\caption{Evaluation performance summary for 80 UCR datasets}
\label{table:ucr_summary}
\end{table*}

%\vspace{-0.3cm}
\subsection{Experiment 4: Mouse dataset experiment}
\label{ssec:real_world_data}
In this section, we compare all the methods on the real-world mouse dynamics dataset.
% (see \Cref{sec:data_utility} to access the data). 
This dataset was collected by drawing four types of gestures with the mouse device: $8$, $\&$, $\%$ and $\star$.
The mouse location coordinates (x and y) were recorded with a sampling frequency of $20$Hz. 
Before the acquisition process every user (seven in total) was instructed by providing an illustration with the trajectories of the four gestures which he or she had to replicate for eight times in turn (every user performed $32$ gestures in total). 
In between of every two gestures the user was free to perform any random mouse movements. 
There were $224$ gestures collected in total from all the users together with random mouse movements forming a data stream.
The resulting data stream was labeled for testing purposes but kept unsegmented. 
To assess the performance of the proposed DNRTPM and the other methods every labeled gesture subsequence was used as the query sequence for a top $k  (k=56, \text{the amount of every gesture in data stream})$ query task resulting in $224$ top $k$ queries. The measured recall, AoD and DTW distance metrics for the all methods are provided in \Cref{table:mouse_dynamics}.

\Cref{table:mouse_dynamics} again shows the superior subsequence discovery performance of the proposed DNRTPM than the other methods in a real-world scenario. DNRTPM obtains significantly higher recall as well as provides quality retrieval of subsequences according to DTW distance metric.

\begin{table}[htb]
\resizebox{\linewidth}{!}{
\begin{tabular}{|l|l|l|l|l|l|l|}
\hline
\multicolumn{2}{|l|}{ } & ISPRING & NSPRING &  \begin{tabular}[c]{@{}l@{}}UCR-\\DTW\end{tabular} & \begin{tabular}[c]{@{}l@{}}UCR-\\US\end{tabular} & DNRTPM \\ \hline
\multirow{2}{*}{\begin{tabular}[c]{@{}l@{}}X\\ axis\end{tabular}} & Recall & 0.58 & 0.54 & 0.58 & 0.32 & \textbf{0.74} \\ \cline{2-7} 
 %& AoD & 0.83 & 0.84 & 0.80 & \textbf{0.86} & \textbf{0.86} \\ \cline{2-7} 
 & \begin{tabular}[c]{@{}l@{}}DTW\\  distance\end{tabular} & 13.76 & 15.01 & 16.36 & 20.94& \textbf{12.44}\\ \hline
 
\multirow{2}{*}{\begin{tabular}[c]{@{}l@{}}Y\\ axis\end{tabular}} & Recall & 0.52 & 0.50 & 0.52 &0.35 & \textbf{0.62} \\ \cline{2-7} 
% & AoD & 0.80 & 0.82 & 0.81 &\textbf{0.87} & 0.83 \\ \cline{2-7} 
 & \begin{tabular}[c]{@{}l@{}}DTW\\  distance\end{tabular}  & 10.64 & 11.28 & 12.62 &15.11 & \textbf{10.28} \\ \hline
% \multirow{2}{*}{\begin{tabular}[c]{@{}l@{}}Average over \\ both axises\end{tabular}} & Recall & 0.529 & 0.507 & 0.544 & \textbf{0.665} \\ \cline{2-6} 
% & AoD & 0.809 & 0.829 & 0.806 & \textbf{0.841} \\ \hline
\end{tabular}}
\caption{The recall, AoD and DTW distance measures averaged over $224$ top $k$ $(k=56)$ queries on the continuous mouse dynamics data stream}
\label{table:mouse_dynamics}
\end{table}

%\vspace{-0.15cm}

\subsection{Experiment 5: Scalability and time delay}
\label{ssec:Scalability}
\begin{figure}[htb]
\begin{minipage}[b]{1.0\linewidth}
  \centering
  \centerline{\epsfig{figure=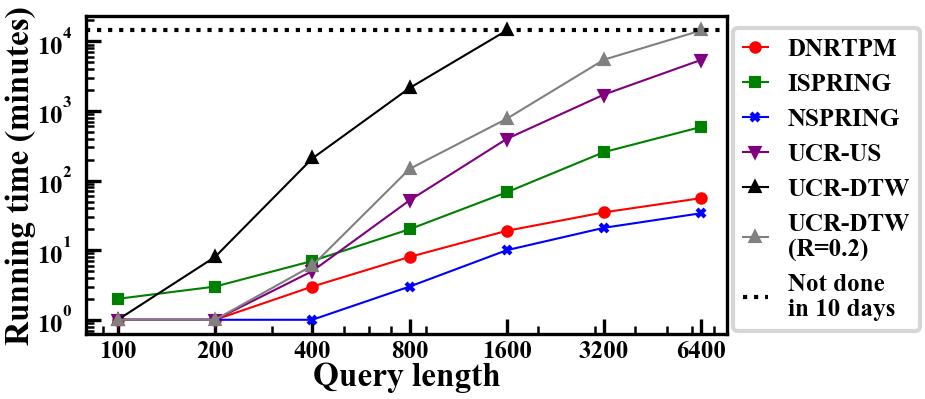, width=8.5cm}}
%  \vspace{2.0cm}
  \caption{Running time for all methods.}
  \label{fig:running_time}
\end{minipage}
\end{figure}

\begin{figure}[htb]
\begin{minipage}[b]{1.0\linewidth}
  \centering
  \centerline{\epsfig{figure=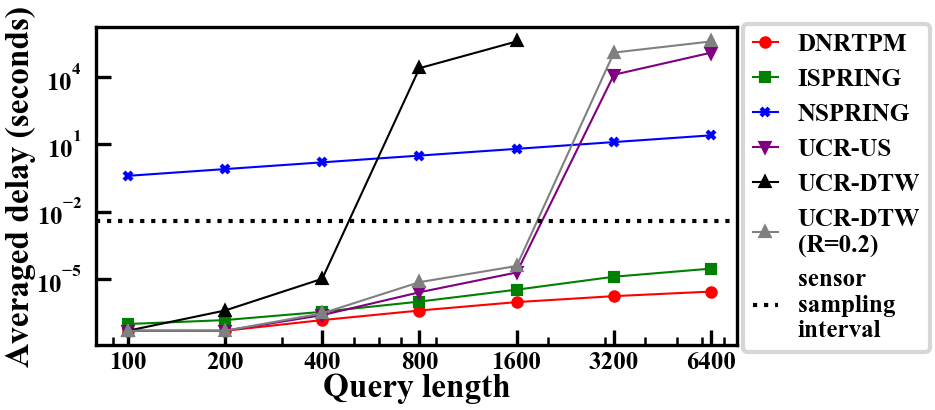, width=8.5cm}}
%  \vspace{2.0cm}
  \caption{Averaged time delay for all methods.}
  \label{fig:delays}
\end{minipage}
\end{figure}

In this experiment, we test the scalability and the time delay of all methods by performing top $k=50$ query task on a ECG dataset~\cite{DTW_million}. The ECG dataset has $n=20,140,000$ data points and was recorded with a sensor samping interval of $t_s = 0.004$ seconds. The running time for each query length is the averaged running time of ten top $k$ queries using randomly selected query sequences. For UCR-DTW we additionally apply a Sakoe-Chiba Band ($R=0.2$) constraining the warping path to speed up calculation. The running time for all methods is shown in \Cref{fig:running_time}. Time delay is defined as the difference between the time when a data point arrives and the time when its processing is finished. The processing time per data point $t_p$ averaged over all data points is obtained by dividing the running time by the number of data points ($n$). For ISPRING, UCR-US, UCR-DTW and DNRTPM, the averaged delay $\Delta t$ can be estimated by: 1) when $dt_p<dt_s$ the averaged time delay equals to the averaged time of processing a data point ($\Delta t = dt_p$); 2) when $dt_p \geq dt_s$ the processing of a data point is still not finished when the next data point arrives, so the time delay is accumulated for each new data point. In this case, the averaged time delay can be estimated by $\Delta t = \frac{n(dt_p-dt_s)}{2}+dt_s$. NSPRING processes each data point when $m$ (length of the query) subsequent data points are available~\cite{gong_si_fong_mohammed_2014}, so its averaged time delay can be obtained by: $\Delta t = mdt_s+dt_p$ when $dt_p<dt_s$ and $\Delta t = \frac{n(dt_p-dt_s)}{2}+(m+1)dt_s$ when $dt_p \geq dt_s$. The averaged time delays for all methods are shown in \Cref{fig:delays}.

%\vspace{-0.1cm}

\Cref{fig:running_time} demonstrates scalability properties of DNRTPM. The running time of UCR-DTW, UCR-US and ISPRING is significantly higher than that of NSPRING and the proposed DNRTPM, limiting their applications especially when the query sequence is very long. \Cref{fig:delays} reveals that the time delay for DNRTPM is the smallest among all methods and is bellow the sensor sampling interval so that it is able to respond in real-time. The time delay of NSPRING, UCR-US and UCRDTW can exceed the sensor sampling interval which might limit their practical use in real-time monitoring scenarios. 
%Note should be taken, that the running speed of UCR-DTW is significantly lower than the one of DNRTPM which might limit its practical applications. 
Note should be taken, that although ISPRING also provides low averaged time delay, its worst case time complexity is $O(m^2)$ per time tick which happens when the minimum or maximum values in its monitoring window change~\cite{ISPRING} and can cause unexpected time delay at certain time ticks. To the contrast, the proposed DNRTPM guarantees time complexity of $O(m)$ per time tick. 

\section{Conclusion}
\label{sec:conclusions}
In this paper, we introduced a real-time pattern matching approach that is based on the dynamic z-normalization scheme and is robust to time and amplitude distortions of different degree. We proved that the introduced dynamic z-normalization provides similar results to the traditional z-normalization performed on the proper (but in practice unknown) window. We demonstrated that the proposed pattern matching method provides high operational performance on both synthetic and real-world scenarios outperforming the other state-of-the-art pattern matching methods.

\bibliography{refs}


\begin{thebibliography}{30}

%%% the Web address.

\ifx \showCODEN    \undefined \def \showCODEN     #1{\unskip}     \fi
\ifx \showDOI      \undefined \def \showDOI       #1{#1}\fi
\ifx \showISBNx    \undefined \def \showISBNx     #1{\unskip}     \fi
\ifx \showISBNxiii \undefined \def \showISBNxiii  #1{\unskip}     \fi
\ifx \showISSN     \undefined \def \showISSN      #1{\unskip}     \fi
\ifx \showLCCN     \undefined \def \showLCCN      #1{\unskip}     \fi
\ifx \shownote     \undefined \def \shownote      #1{#1}          \fi
\ifx \showarticletitle \undefined \def \showarticletitle #1{#1}   \fi
\ifx \showURL      \undefined \def \showURL       {\relax}        \fi
% invisible to TeX
\providecommand\bibfield[2]{#2}
\providecommand\bibinfo[2]{#2}
\providecommand\natexlab[1]{#1}
\providecommand\showeprint[2][]{arXiv:#2}

\bibitem[\protect\citeauthoryear{A.W.Fu, E.Keogh, L.Y.Lau, C.A.Ratanamahatana,
  and R.C.Wong}{A.W.Fu et~al\mbox{.}}{2008}]%
        {fu2008scaling}
\bibfield{author}{\bibinfo{person}{A.W.Fu}, \bibinfo{person}{E.Keogh},
  \bibinfo{person}{L.Y.Lau}, \bibinfo{person}{C.A.Ratanamahatana}, {and}
  \bibinfo{person}{R.C.Wong}.} \bibinfo{year}{2008}\natexlab{}.
\newblock \showarticletitle{Scaling and time warping in time series querying}.
\newblock \bibinfo{journal}{\emph{The International Journal on Very Large Data
  Bases}} \bibinfo{volume}{17}, \bibinfo{number}{4} (\bibinfo{year}{2008}),
  \bibinfo{pages}{899--921}.
\newblock


\bibitem[\protect\citeauthoryear{A.Yang, A.Giani, R.Giannatonio, K.Gilani,
  et~al\mbox{.}}{A.Yang et~al\mbox{.}}{2009}]%
        {yang2009distributed}
\bibfield{author}{\bibinfo{person}{A.Yang}, \bibinfo{person}{A.Giani},
  \bibinfo{person}{R.Giannatonio}, \bibinfo{person}{K.Gilani}, {et~al\mbox{.}}}
  \bibinfo{year}{2009}\natexlab{}.
\newblock \showarticletitle{Distributed human action recognition via wearable
  motion sensor networks}.
\newblock \bibinfo{journal}{\emph{Journal of Ambient Intelligence and Smart
  Environments}} \bibinfo{volume}{1}, \bibinfo{number}{2}
  (\bibinfo{year}{2009}), \bibinfo{pages}{103--115}.
\newblock


\bibitem[\protect\citeauthoryear{B.C.Giao and D.T.Anh}{B.C.Giao and
  D.T.Anh}{2016}]%
        {ISPRING}
\bibfield{author}{\bibinfo{person}{B.C.Giao} {and} \bibinfo{person}{D.T.Anh}.}
  \bibinfo{year}{2016}\natexlab{}.
\newblock \showarticletitle{Improving spring method in similarity search over
  time-series streams by data normalization}. In
  \bibinfo{booktitle}{\emph{International Conference on Nature of Computation
  and Communication}}. Springer, \bibinfo{pages}{189--202}.
\newblock


\bibitem[\protect\citeauthoryear{B.Hu, Y.Chen, and E.Keogh}{B.Hu
  et~al\mbox{.}}{2013}]%
        {hu2013time}
\bibfield{author}{\bibinfo{person}{B.Hu}, \bibinfo{person}{Y.Chen}, {and}
  \bibinfo{person}{E.Keogh}.} \bibinfo{year}{2013}\natexlab{}.
\newblock \showarticletitle{Time series classification under more realistic
  assumptions}. In \bibinfo{booktitle}{\emph{Proceedings of the 2013 SIAM
  International Conference on Data Mining}}. SIAM, \bibinfo{pages}{578--586}.
\newblock


\bibitem[\protect\citeauthoryear{C.A.Ratanamahatana and
  E.Keogh}{C.A.Ratanamahatana and E.Keogh}{2004}]%
        {ratanamahatana2004making}
\bibfield{author}{\bibinfo{person}{C.A.Ratanamahatana} {and}
  \bibinfo{person}{E.Keogh}.} \bibinfo{year}{2004}\natexlab{}.
\newblock \showarticletitle{Making time-series classification more accurate
  using learned constraints}. In \bibinfo{booktitle}{\emph{Proceedings of the
  2004 SIAM International Conference on Data Mining}}. SIAM,
  \bibinfo{pages}{11--22}.
\newblock


\bibitem[\protect\citeauthoryear{CMU}{CMU}{[n. d.]}]%
        {CMU_database}
CMU \bibinfo{year}{[n. d.]}\natexlab{}.
\newblock \bibinfo{title}{Graphics Lab Motion Capture Database}.
\newblock
\newblock
\urldef\tempurl%
\url{mocap.cs.cmu.edu}
\showURL{%
Retrieved October 23, 2018 from \tempurl}


\bibitem[\protect\citeauthoryear{D.J.Berndt and J.Clifford}{D.J.Berndt and
  J.Clifford}{1994}]%
        {berndt1994using}
\bibfield{author}{\bibinfo{person}{D.J.Berndt} {and}
  \bibinfo{person}{J.Clifford}.} \bibinfo{year}{1994}\natexlab{}.
\newblock \showarticletitle{Using dynamic time warping to find patterns in time
  series.}. In \bibinfo{booktitle}{\emph{KDD workshop}},
  Vol.~\bibinfo{volume}{10}. Seattle, WA, \bibinfo{pages}{359--370}.
\newblock


\bibitem[\protect\citeauthoryear{E.Keogh and S.Kasetty}{E.Keogh and
  S.Kasetty}{2003}]%
        {keogh2003need}
\bibfield{author}{\bibinfo{person}{E.Keogh} {and} \bibinfo{person}{S.Kasetty}.}
  \bibinfo{year}{2003}\natexlab{}.
\newblock \showarticletitle{On the need for time series data mining benchmarks:
  a survey and empirical demonstration}.
\newblock \bibinfo{journal}{\emph{Data Mining and knowledge discovery}}
  \bibinfo{volume}{7}, \bibinfo{number}{4} (\bibinfo{year}{2003}),
  \bibinfo{pages}{349--371}.
\newblock


\bibitem[\protect\citeauthoryear{E.Keogh, T.Palpanas, V.Zordan, D.Gunopulos,
  and M.Cardle}{E.Keogh et~al\mbox{.}}{2004}]%
        {keogh2004indexing}
\bibfield{author}{\bibinfo{person}{E.Keogh}, \bibinfo{person}{T.Palpanas},
  \bibinfo{person}{V.Zordan}, \bibinfo{person}{D.Gunopulos}, {and}
  \bibinfo{person}{M.Cardle}.} \bibinfo{year}{2004}\natexlab{}.
\newblock \showarticletitle{Indexing large human-motion databases}. In
  \bibinfo{booktitle}{\emph{Proceedings of the Thirtieth international
  conference on Very large data bases-Volume 30}}. VLDB Endowment,
  \bibinfo{pages}{780--791}.
\newblock


\bibitem[\protect\citeauthoryear{E.Ogasawara, L.C.Martinez, D.Oliveira,
  G.Zimbr{\~a}o, G.L.Pappa, and M.Mattoso}{E.Ogasawara et~al\mbox{.}}{2010}]%
        {ogasawara2010adaptive}
\bibfield{author}{\bibinfo{person}{E.Ogasawara},
  \bibinfo{person}{L.C.Martinez}, \bibinfo{person}{D.Oliveira},
  \bibinfo{person}{G.Zimbr{\~a}o}, \bibinfo{person}{G.L.Pappa}, {and}
  \bibinfo{person}{M.Mattoso}.} \bibinfo{year}{2010}\natexlab{}.
\newblock \showarticletitle{Adaptive normalization: A novel data normalization
  approach for non-stationary time series}. In \bibinfo{booktitle}{\emph{The
  2010 International Joint Conference on Neural Networks (IJCNN)}}. IEEE,
  \bibinfo{pages}{1--8}.
\newblock


\bibitem[\protect\citeauthoryear{H.A.Dau, E.Keogh, K.Kamgar, C.M.Yeh, Y.Zhu,
  S.Gharghabi, C.A.Ratanamahatana, Yanping, B.Hu, N.Begum, A.Bagnall, A.Mueen,
  and G.Batista}{H.A.Dau et~al\mbox{.}}{2018}]%
        {UCRArchive2018}
\bibfield{author}{\bibinfo{person}{H.A.Dau}, \bibinfo{person}{E.Keogh},
  \bibinfo{person}{K.Kamgar}, \bibinfo{person}{C.M.Yeh},
  \bibinfo{person}{Y.Zhu}, \bibinfo{person}{S.Gharghabi},
  \bibinfo{person}{C.A.Ratanamahatana}, \bibinfo{person}{Yanping},
  \bibinfo{person}{B.Hu}, \bibinfo{person}{N.Begum},
  \bibinfo{person}{A.Bagnall}, \bibinfo{person}{A.Mueen}, {and}
  \bibinfo{person}{G.Batista}.} \bibinfo{year}{2018}\natexlab{}.
\newblock \bibinfo{title}{The {UCR} Time Series Classification Archive}.
\newblock
\newblock
\newblock
\shownote{\url{https://www.cs.ucr.edu/~eamonn/time_series_data_2018/}.}


\bibitem[\protect\citeauthoryear{H.Ding, G.Trajcevski, P.Scheuermann, X.Wang,
  and E.Keogh}{H.Ding et~al\mbox{.}}{2008}]%
        {ding_trajcevski_scheuermann_wang_keogh_2008}
\bibfield{author}{\bibinfo{person}{H.Ding}, \bibinfo{person}{G.Trajcevski},
  \bibinfo{person}{P.Scheuermann}, \bibinfo{person}{X.Wang}, {and}
  \bibinfo{person}{E.Keogh}.} \bibinfo{year}{2008}\natexlab{}.
\newblock \showarticletitle{Querying and mining of time series data:
  experimental comparison of representations and distance measures.}. In
  \bibinfo{booktitle}{\emph{Proceedings of the VLDB Endowment}},
  Vol.~\bibinfo{volume}{1}. \bibinfo{pages}{1542–1552}.
\newblock


\bibitem[\protect\citeauthoryear{J.Aach and G.M.Church}{J.Aach and
  G.M.Church}{2001}]%
        {aach2001aligning}
\bibfield{author}{\bibinfo{person}{J.Aach} {and} \bibinfo{person}{G.M.Church}.}
  \bibinfo{year}{2001}\natexlab{}.
\newblock \showarticletitle{Aligning gene expression time series with time
  warping algorithms}.
\newblock \bibinfo{journal}{\emph{Bioinformatics}} \bibinfo{volume}{17},
  \bibinfo{number}{6} (\bibinfo{year}{2001}), \bibinfo{pages}{495--508}.
\newblock


\bibitem[\protect\citeauthoryear{J.Kruskal and M.Liberman}{J.Kruskal and
  M.Liberman}{1983}]%
        {kruskalsymmetric}
\bibfield{author}{\bibinfo{person}{J.Kruskal} {and}
  \bibinfo{person}{M.Liberman}.} \bibinfo{year}{1983}\natexlab{}.
\newblock \bibinfo{title}{The Symmetric Time-Warping Problem: From Continuous
  to Discrete.}
\newblock
\newblock


\bibitem[\protect\citeauthoryear{M.M{\"u}ller}{M.M{\"u}ller}{2007}]%
        {muller2007dynamic}
\bibfield{author}{\bibinfo{person}{M.M{\"u}ller}.}
  \bibinfo{year}{2007}\natexlab{}.
\newblock \showarticletitle{Dynamic time warping}.
\newblock \bibinfo{journal}{\emph{Information retrieval for music and motion}}
  (\bibinfo{year}{2007}), \bibinfo{pages}{69--84}.
\newblock


\bibitem[\protect\citeauthoryear{PAMAP}{PAMAP}{2012}]%
        {PAMAP}
PAMAP \bibinfo{year}{2012}\natexlab{}.
\newblock \bibinfo{title}{Physical Activity Monitoring for Aging People}.
\newblock
\newblock
\urldef\tempurl%
\url{www.pamap.org/demo.html}
\showURL{%
Retrieved October 23, 2018 from \tempurl}


\bibitem[\protect\citeauthoryear{P.Koch, W.Konen, and K.Hein}{P.Koch
  et~al\mbox{.}}{2010}]%
        {koch2010gesture}
\bibfield{author}{\bibinfo{person}{P.Koch}, \bibinfo{person}{W.Konen}, {and}
  \bibinfo{person}{K.Hein}.} \bibinfo{year}{2010}\natexlab{}.
\newblock \showarticletitle{Gesture recognition on few training data using slow
  feature analysis and parametric bootstrap}. In
  \bibinfo{booktitle}{\emph{Neural Networks (IJCNN), The 2010 International
  Joint Conference on}}. Citeseer, \bibinfo{pages}{1--8}.
\newblock


\bibitem[\protect\citeauthoryear{P.N.Tan, M.Steinbach, and V.Kumar}{P.N.Tan
  et~al\mbox{.}}{2014}]%
        {tan_steinbach_kumar_2014}
\bibfield{author}{\bibinfo{person}{P.N.Tan}, \bibinfo{person}{M.Steinbach},
  {and} \bibinfo{person}{V.Kumar}.} \bibinfo{year}{2014}\natexlab{}.
\newblock \bibinfo{booktitle}{\emph{Introduction to data mining}}.
\newblock \bibinfo{publisher}{Pearson Addison Wesley}.
\newblock


\bibitem[\protect\citeauthoryear{T.Argyros and C.Ermopoulos}{T.Argyros and
  C.Ermopoulos}{2003}]%
        {argyros2003efficient}
\bibfield{author}{\bibinfo{person}{T.Argyros} {and}
  \bibinfo{person}{C.Ermopoulos}.} \bibinfo{year}{2003}\natexlab{}.
\newblock \showarticletitle{Efficient subsequence matching in time series
  databases under time and amplitude transformations}. In
  \bibinfo{booktitle}{\emph{Data Mining, 2003. ICDM 2003. Third IEEE
  International Conference on}}. IEEE, \bibinfo{pages}{481--484}.
\newblock


\bibitem[\protect\citeauthoryear{T.Fu}{T.Fu}{2011}]%
        {fu2011review}
\bibfield{author}{\bibinfo{person}{T.Fu}.} \bibinfo{year}{2011}\natexlab{}.
\newblock \showarticletitle{A review on time series data mining}.
\newblock \bibinfo{journal}{\emph{Engineering Applications of Artificial
  Intelligence}} \bibinfo{volume}{24}, \bibinfo{number}{1}
  (\bibinfo{year}{2011}), \bibinfo{pages}{164--181}.
\newblock


\bibitem[\protect\citeauthoryear{T.M.Rath and R.Manmatha}{T.M.Rath and
  R.Manmatha}{2003}]%
        {rath2003word}
\bibfield{author}{\bibinfo{person}{T.M.Rath} {and}
  \bibinfo{person}{R.Manmatha}.} \bibinfo{year}{2003}\natexlab{}.
\newblock \showarticletitle{Word image matching using dynamic time warping}. In
  \bibinfo{booktitle}{\emph{Computer Vision and Pattern Recognition, 2003.
  Proceedings. 2003 IEEE Computer Society Conference on}},
  Vol.~\bibinfo{volume}{2}. IEEE, \bibinfo{pages}{II--II}.
\newblock


\bibitem[\protect\citeauthoryear{T.Rakthanmanon, B.Campana, A.Mueen, G.Batista,
  B.Westover, Q.Zhu, J.Zakaria, and E.Keogh}{T.Rakthanmanon
  et~al\mbox{.}}{2012}]%
        {DTW_million}
\bibfield{author}{\bibinfo{person}{T.Rakthanmanon},
  \bibinfo{person}{B.Campana}, \bibinfo{person}{A.Mueen},
  \bibinfo{person}{G.Batista}, \bibinfo{person}{B.Westover},
  \bibinfo{person}{Q.Zhu}, \bibinfo{person}{J.Zakaria}, {and}
  \bibinfo{person}{E.Keogh}.} \bibinfo{year}{2012}\natexlab{}.
\newblock \showarticletitle{Searching and mining trillions of time series
  subsequences under dynamic time warping}. In
  \bibinfo{booktitle}{\emph{Proceedings of the 18th ACM SIGKDD international
  conference on Knowledge discovery and data mining - KDD 12}}.
\newblock


\bibitem[\protect\citeauthoryear{T.Rakthanmanon, B.Campana, A.Mueen, G.Batista,
  B.Westover, Q.Zhu, J.Zakaria, and E.Keogh}{T.Rakthanmanon
  et~al\mbox{.}}{2013}]%
        {DTW_million_journal}
\bibfield{author}{\bibinfo{person}{T.Rakthanmanon},
  \bibinfo{person}{B.Campana}, \bibinfo{person}{A.Mueen},
  \bibinfo{person}{G.Batista}, \bibinfo{person}{B.Westover},
  \bibinfo{person}{Q.Zhu}, \bibinfo{person}{J.Zakaria}, {and}
  \bibinfo{person}{E.Keogh}.} \bibinfo{year}{2013}\natexlab{}.
\newblock \showarticletitle{Addressing big data time series: Mining trillions
  of time series subsequences under dynamic time warping}.
\newblock \bibinfo{journal}{\emph{ACM Transactions on Knowledge Discovery from
  Data (TKDD)}} \bibinfo{volume}{7}, \bibinfo{number}{3}
  (\bibinfo{year}{2013}), \bibinfo{pages}{10}.
\newblock


\bibitem[\protect\citeauthoryear{V.Niennattrakul, D.Wanichsan, and
  C.A.Ratanamahatana}{V.Niennattrakul et~al\mbox{.}}{2010}]%
        {niennattrakul_wanichsan_ratanamahatana_2010}
\bibfield{author}{\bibinfo{person}{V.Niennattrakul},
  \bibinfo{person}{D.Wanichsan}, {and} \bibinfo{person}{C.A.Ratanamahatana}.}
  \bibinfo{year}{2010}\natexlab{}.
\newblock \showarticletitle{Accurate Subsequence Matching on Data Stream under
  Time Warping Distance}.
\newblock \bibinfo{journal}{\emph{New Frontiers in Applied Data Mining Lecture
  Notes in Computer Science}} (\bibinfo{year}{2010}),
  \bibinfo{pages}{156–167}.
\newblock


\bibitem[\protect\citeauthoryear{Wu}{Wu}{2020}]%
        {wurenzhi2020Apr}
\bibfield{author}{\bibinfo{person}{Renzhi Wu}.}
  \bibinfo{year}{2020}\natexlab{}.
\newblock \bibinfo{title}{{DNRTPM}}.
\newblock
\newblock
\urldef\tempurl%
\url{https://github.com/wurenzhi/DNRTPM}
\showURL{%
\tempurl}
\newblock
\shownote{[Online; accessed 8. Apr. 2020].}


\bibitem[\protect\citeauthoryear{X.Gong, S.Fong, and Y.Si}{X.Gong
  et~al\mbox{.}}{2018}]%
        {gong_fong_si_2018}
\bibfield{author}{\bibinfo{person}{X.Gong}, \bibinfo{person}{S.Fong}, {and}
  \bibinfo{person}{Y.Si}.} \bibinfo{year}{2018}\natexlab{}.
\newblock \showarticletitle{Fast multi-subsequence monitoring on streaming
  time-series based on Forward-propagation}.
\newblock \bibinfo{journal}{\emph{Information Sciences}}  \bibinfo{volume}{450}
  (\bibinfo{year}{2018}), \bibinfo{pages}{73–88}.
\newblock


\bibitem[\protect\citeauthoryear{X.Gong, Y.Si, S.Fong, and S.Mohammed}{X.Gong
  et~al\mbox{.}}{2014}]%
        {gong_si_fong_mohammed_2014}
\bibfield{author}{\bibinfo{person}{X.Gong}, \bibinfo{person}{Y.Si},
  \bibinfo{person}{S.Fong}, {and} \bibinfo{person}{S.Mohammed}.}
  \bibinfo{year}{2014}\natexlab{}.
\newblock \showarticletitle{{NSPRING}: Normalization-supported SPRING for
  subsequence matching on time series streams}. In
  \bibinfo{booktitle}{\emph{2014 IEEE 15th International Symposium on
  Computational Intelligence and Informatics (CINTI)}}.
\newblock


\bibitem[\protect\citeauthoryear{Y.Sakurai, C.Faloutsos, and
  M.Yamamuro}{Y.Sakurai et~al\mbox{.}}{2007}]%
        {sakurai_faloutsos_yamamuro_2007}
\bibfield{author}{\bibinfo{person}{Y.Sakurai}, \bibinfo{person}{C.Faloutsos},
  {and} \bibinfo{person}{M.Yamamuro}.} \bibinfo{year}{2007}\natexlab{}.
\newblock \showarticletitle{Stream Monitoring under the Time Warping Distance}.
  In \bibinfo{booktitle}{\emph{2007 IEEE 23rd International Conference on Data
  Engineering}}.
\newblock


\bibitem[\protect\citeauthoryear{Y.Shen, Y.Chen, E.Keogh, and H.Jin}{Y.Shen
  et~al\mbox{.}}{2017}]%
        {shen2017searching}
\bibfield{author}{\bibinfo{person}{Y.Shen}, \bibinfo{person}{Y.Chen},
  \bibinfo{person}{E.Keogh}, {and} \bibinfo{person}{H.Jin}.}
  \bibinfo{year}{2017}\natexlab{}.
\newblock \showarticletitle{Searching time series with invariance to large
  amounts of uniform scaling}. In \bibinfo{booktitle}{\emph{Data Engineering
  (ICDE), 2017 IEEE 33rd International Conference on}}. IEEE,
  \bibinfo{pages}{111--114}.
\newblock


\bibitem[\protect\citeauthoryear{Y.Shen, Y.Chen, E.Keogh, and H.Jin}{Y.Shen
  et~al\mbox{.}}{2018}]%
        {shen_chen_keogh_jin_2018}
\bibfield{author}{\bibinfo{person}{Y.Shen}, \bibinfo{person}{Y.Chen},
  \bibinfo{person}{E.Keogh}, {and} \bibinfo{person}{H.Jin}.}
  \bibinfo{year}{2018}\natexlab{}.
\newblock \showarticletitle{Accelerating Time Series Searching with Large
  Uniform Scaling}. In \bibinfo{booktitle}{\emph{Proceedings of the 2018 SIAM
  International Conference on Data Mining}}. SIAM, \bibinfo{pages}{234--242}.
\newblock


\end{thebibliography}
\bibliographystyle{ACM-Reference-Format}
\newpage
\clearpage
\appendix
\section{Proof}
\label{sec:proof_invariants}
Proof for \Cref{eq:invariants}. For a continuous function $f(x)$ defined on a continuous interval $x \in [x_l,x_u]$, the prefix normalization is $f(x)' = \frac{f(x)-\mu_x}{\sigma_x}$
%\begin{equation}
%\label{eq:prefix_norm_proof}
%\begin{aligned}
%f(x)' =& \frac{f(x)-\mu_x}{\sigma_x}
%\end{aligned}
%\end{equation}
where $\mu_x = \frac{\int_{x_l}^{x} f(t) dt}{x-x_l}$ and $\sigma_x = \sqrt{\frac{\int_{x_l}^{x} f(t)^2 dt}{x-x_l} - \mu_x^2}$.
%\begin{equation}
%\label{eq:prefix_mu_proof}
%\mu_x = \frac{1}{x-x_l}\int_{x_l}^{x} f(t) dt;\ \ \sigma_x = \sqrt{(\frac{1}{x-x_l}\int_{x_l}^{x} f(t)^2 dt - \mu_x^2)}
%\end{equation}
The amplification and shift factors are $\eta_x = \frac{\sigma_x}{\sigma_{x_u}}$ and $\delta_x = \frac{\mu_x-\mu_{x_u}}{\sigma_{x_u}}$.
%\begin{equation}
%\label{eq:factors_proof}
%\begin{aligned}
%\eta_x = \frac{\sigma_x}{\sigma_{x_u}},\quad \delta_x = \frac{\mu_x-\mu_{x_u}}{\sigma_{x_u}}\\
%\end{aligned}
%\end{equation}
For function $g(x) = C_2 f(C_1 (x+C_0))+C_3$, its prefix mean is:
\begin{equation}
\label{eq:prefix_mu_proof}
\begin{aligned}
\mu_{x'}' =& \frac{\int_{x_l'}^{x'} g(t) dt}{x'-x_l'}= \frac{C_2\int_{\frac{x_l}{C_1}-C_0}^{x'} f(C_1 (t+C_0))dt}{x' - \frac{x_l}{C_1}+C_0} + C_3\\
=& \frac{C_2\int_{x_l}^{(x'+C_0)C_1} f(v)dv}{C_1 x' - x_l+ C_0 C_1} + C_3 \ \ \ \ (\text{let} \ v = C_1 (t+C_0))
\end{aligned} 
\end{equation}
The prefix standard deviation is:
\begin{equation}
\label{eq:prefix_sigma_proof}
\begin{aligned}
\sigma_{x'}' =& \sqrt{\frac{\int_{x_l'}^{x'} g(t)^2 dt}{x'-x_l'} - \mu_{x'}^2}\\
=& \sqrt{\frac{\int_{x_l'}^{x'} (C_2 f(C_1 (t+C_0))+C_3)^2 dt}{x'-x_l'} - \mu_{x'}^2}\\
=& \sqrt{\frac{\int_{x_l}^{(x'+C_0)C_1} (C_2 f(v)+C_3)^2 dv}{C_1 x' - x_l+ C_0 C_1} - \mu_{x'}^2} \ \ \ \ (\text{let} \ v = C_1 (t+C_0))\\
=& \sqrt{\frac{\int_{x_l}^{(x'+C_0)C_1} C_2^2 f(v)^2dv - \frac{C_2^2}{C_1 x' - x_l+ C_0 C_1}(\int_{x_l}^{(x'+C_0)C_1} f(v)dv)^2}{C_1 x' - x_l+ C_0 C_1}}\\
\end{aligned}
\end{equation}

Under the condition that $\frac{x-x_l}{x_u-x_l} = \frac{x'-x_l'}{x_u'- x_l'}$:
\begin{equation}
\label{eq:x_x_proof}
x' = \frac{x}{C_1} - C_0
\end{equation}
Substitute \Cref{eq:x_x_proof} into \Cref{eq:prefix_mu_proof} and \Cref{eq:prefix_sigma_proof}:
\begin{small}
\begin{equation}
\begin{aligned}
\mu_{x'}' = & \frac{C_2\int_{x_l}^{(x'+C_0)C_1} f(v)dv}{C_1 x' - x_l+ C_0 C_1} + C_3 = \frac{C_2\int_{x_l}^{x} f(v)dv}{x - x_l} + C_3=C_2 \mu_x + C_3\\
\sigma_{x'}' = &
\sqrt{\frac{\int_{x_l}^{(x'+C_0)C_1} C_2^2 f(v)^2dv - \frac{C_2^2}{C_1 x' - x_l+ C_0 C_1}(\int_{x_l}^{(x'+C_0)C_1} f(v)dv)^2}{C_1 x' - x_l+ C_0 C_1}}\\
=&C_2 \sqrt{\frac{\int_{x_l}^{x} f(v)^2dv - \frac{1}{x - x_l}(\int_{x_l}^{x} f(v)dv)^2}{x- x_l}}=C_2 \sigma_x\\
\end{aligned}
\end{equation}
\end{small}
Therefore:
\begin{equation}
\begin{aligned}
\eta_{x'}' =& \frac{\sigma_{x'}'}{\sigma_{x_u'}'}
= \frac{C_2\sigma_{x}}{C_2 \sigma_{x_u}}
= \eta_{x}\\
\delta_{x'}' = & \frac{\mu_{x'}'-\mu_{x_u'}'}{\sigma_{x'}'}
=\frac{C_2 \mu_{x}+C_3-C_2 \mu_{x_u} - C_3}{C_2 \sigma_{x}}
=&\delta_x\\
\end{aligned}
\end{equation}

\section{DNRTPM pseudocode}
\label{sec:pseudocode}
The pseudocode of the proposed DNRTPM is summarized in \Cref{alg:DNRTPM}:
\begin{algorithm}[!htb]
\caption{DNRTPM}
\label{alg:DNRTPM}
\SetAlgoLined
\KwIn{A new value $s_t$ at time-tick $t$}
\KwOut{Matched subsequence $S_{i,t}$ if any}
\textbf{Initialization:} Before the first time tick, initialize $\text{PS}$ and $\text{PSS}$ as empty deque; obtain the prefix normalization and the amplification factor of the query sequence $Q$ by \cref{eq:prefix_norm} and \cref{eq:enlargement_factor}\\
\textbf{end initialization}\\
Compute $\text{ps}_t$, $\text{pss}_t$ by \Cref{eq:prefix_sums} and append them to the end of $\text{PS}$ and $\text{PSS}$;\\
\For{$k\gets 0$ \KwTo $m-1$}{
Compute $D_{t,k}$ and $B_{t,k}$ by \Cref{eq:spring_D} and \Cref{eq:spring_B};
}

Remove $\{\text{ps}_{(B(t-1,min)-1)},\dots,\text{ps}_{(B(t,min)-2)}\}$ and $\{\text{pss}_{(B(t-1,min)-1)},\dots,\text{pss}_{(B(t,min)-2)}\}$ from the beginning of $\text{PS}$ and $\text{PSS}$; 

\If{$D_{min} < \epsilon$}{
	\If{$\forall_k,\  D(t,k) \geq D_{min} \lor B(t,k) > t_e$ }{
	output $D_{min}$, $t_s$, $t_e$;\\\label{alg_line:output}
	$D_{min} = +\infty$;\\
	\For{$k\gets 0$ \KwTo $m-1$}{
      \If{$B(t,k) \leq t_e$}{
      	$D(t,k) = +\infty$;\\
      }
	}
	}
	}
	
\If{$D(t,m-1) < \epsilon$}{\label{alg_line:real_cond}
\If{$D(t,m-1) < D_{min}$}
{
$D_{min} = D(t,m-1)$;\\
$t_s = B(t,m-1)$;\\
$t_e = t$;\\
}
}    
\For{$k\gets 0$ \KwTo $m-1$}{
$D(t-1,k) = D(t,k)$;\\
$B(t-1,k) = B(t,k)$;\\
}
\end{algorithm}

%\section{Source code and data}
%\label{sec:data_utility}
%The data sets, the source code of the proposed and the state-of-the-art methods, as well as all the utility functions that were used in the experiments can be accessed from: \url{https://www.dropbox.com/s/p7enk7layzf2yw4/time_series_pattern_matching.zip?dl=0}

\end{document}